\documentstyle[aps,multicol,eqsecnum,psfig]{revtex}
\begin{document}
\twocolumn[\hsize\textwidth\columnwidth\hsize\csname
@twocolumnfalse\endcsname

\title{Interplay of the Scaling Limit and the Renormalization Group:
Implications for Symmetry Restoration}
\author{Robert M. Konik$^{1,2}$, 
Hubert Saleur$^{3,4}$, and Andreas W. W. Ludwig$^1$.}
\address{
$^1$Department of Physics, UCSB, Santa Barbara, CA 93106}
\address{
$^2$Department of Physics, 
University of Virginia, Charlottesville, VA 22904}
\address{
$^3$Laboratoire de Physique Theorique et Mecanique Statistique, Orsay, France}
\address{$^4$ Department of Physics, University of Southern California,
Los-Angeles, CA 90089-0484}
\date{\today}
\maketitle
\begin{abstract}
Symmetry restoration 
 is usually understood as a renormalization group
induced phenomenon.  
In this context, the issue of whether one-loop RG equations
can be trusted  in predicting symmetry restoration
has recently been  the subject of much debate.
Here we advocate a more pragmatic point of view
and expand the definition of symmetry restoration to encompass all
situations where the physical properties have only
a weak dependence upon an anisotropy in the bare couplings.
Moreover we concentrate on universal properties,
and so take a scaling limit where the 
physics is well described by a field theory.  In this context, we find 
a large variety of models that exhibit, for all practical purposes,
symmetry restoration: even if symmetry is not restored in a strict sense,
physical properties are surprisingly insensitive to the remaining 
anisotropy.  

\hskip 5pt 
Although we have adopted an expanded notion
of symmetry restoration, we nonetheless 
emphasize that the scaling limit also has implications 
for symmetry restoration as a renormalization group induced
phenomenon.  In all the models we considered, the scaling limit
turns out to only permit bare couplings which are
nearly isotropic and {\sl small}. 
Then the one-loop beta-function should contain all the physics
and higher loop orders can be neglected.
We suggest that this 
feature generalizes to more complex models.  We
exhibit a large class of theories with current-current perturbations 
(of which the $SO(8)$ model of interest in two-leg Hubbard ladders/armchair
carbon nanotubes is one) 
where the one-loop
beta-functions indicates symmetry restoration and so argue that these 
results can be trusted within the scaling limit. 
\end{abstract}
\pacs{PACS numbers: ????}

]

\newcommand{\del}{\partial}
\newcommand{\nn}{\nonumber}
\newcommand{\rtc}{\tilde{\rho}_c}
\newcommand{\th}{\theta}
\newcommand{\gcc}{\Gamma_{cc}}
\newcommand{\la}{\lambda}

\setcounter{footnote}{2}

\section{Introduction: Symmetry Restoration}

Under the most prevalent definition, symmetry restoration occurs as
a Hamiltonian is attracted under a renormalization group (RG) flow to
a manifold possessing a higher symmetry than indicated by the original 
bare or microscopic theory.  A recent example in the literature of this
phenomenon is found in the work of Lin et al.\cite{fisher}. 
This work addresses
the low energy behaviour of a two-leg Hubbard ladder (or,
equivalently, the armchair carbon nanotube).  These authors argued
that a Hamiltonian for these systems with 
`generic' (short-range)
interactions flow under an 1-loop RG towards the 
SO(8) Gross-Neveu model, or more
specifically, the set of couplings $\{\lambda_i\}$ flow onto fixed
ratios indicative of SO(8) Gross-Neveu.

Two more examples of symmetry restoration as suggested by the 
1-loop
RG are provided by the anisotropic Kondo model
and the $U(1)$ Thirring model.  The anisotropic Kondo model
is described by
the Hamiltonian
\begin{eqnarray}\label{KH}
H &=& \int dx \sum_\sigma -ic^\dagger_\sigma (x)
\partial_x c_\sigma (x) 
\!+\! {1\over 2}\! \sum_{\sigma \sigma '}c^\dagger_\sigma(0)
\{g_{||}\sigma^z_{\sigma \sigma '} s^z \cr
&& \hskip .5in
+g_\perp(\sigma^x_{\sigma \sigma '} s^x + \sigma^y_{\sigma \sigma '} s^y)\}
c^\dagger_{\sigma '} (0),
\end{eqnarray}
where $c_\sigma (x)$ is a Fermi field of spin, $\sigma$,
while the $U(1)$ Thirring model is given by the Lagrangian
\begin{equation}
{\cal L} = i\bar\psi_\alpha\gamma_u\del^\mu\psi_\alpha + 
{1\over 4}g_\parallel (j_z)^2 + {1\over 4}g_\perp ((j_x)^2+(j_y)^2),
\end{equation}
where $j^a_\mu = \bar\psi_\alpha\gamma_\mu\tau^a_{\alpha\beta}\psi_\beta$,
and $\psi$ is a doublet of Dirac spinors.
In both these models the
1-loop RG
equations read,
\begin{eqnarray}\label{rgeqs}
{dg_\parallel \over dl} &= - c {g_\perp}^2;\cr
{dg_\perp \over dl} &= - c g_\parallel g_\perp ,
\end{eqnarray}
where $c$ is a model dependent constant
\footnote{In the conventions we chose in this paper,
$c>0$ for the $U(1)$ Thirring model and $c<0$ for the
Kondo model.}.
Thus in the regions where the trajectories 
flow to strong coupling, both models are attracted to the
diagonal  $g_\parallel=g_\perp$
(the $SU(2)$ invariant line): more precisely, the ratio 
$|g_\perp / g_\parallel|$ converges to unity.
The natural conclusion one derives from this feature is that the physics
at large distances (large compared with the UV cut-off), or low energies,  
is well described by an isotropic model (usually
an isotropic field theory).

In drawing this conclusion, two immediate difficulties present themselves.
There is the possibility that higher loop orders make the 
diagonal unstable
in which case, of course, the symmetry 
restoration will not in fact occur. 
This possibility will not be relevant for the examples and discussion 
in the paper at hand, although, in general, it is a genuine concern.
Even upon excluding this scenario,
RG flows towards the diagonal do not necessarily indicate
that symmetry restoration takes place. 
Indeed, in all the models of interest here, 
the interactions are of current-current form and
the bare coupling constants are dimensionless. 
Therefore the physics does not have to depend merely on the ratios 
of these coupling constants; it might also depend on
other combinations that do not flow under the RG. 
A case in point is precisely 
the $U(1)$ Thirring model. 
In this model, there is a quantity $\mu$ 
($\mu^2=g_\parallel^2-g_\perp^2$ at lowest order) that
is an RG invariant, 
and on which  physical quantities {\sl do} depend in a non-trivial way.  
Therefore, if initially $g_\parallel\neq g_\perp$,
symmetry restoration as defined above does not ever occur.
Another example is provided by the anisotropic Kondo model.
Although the physics of the fixed point of this model is isotropic,
the physics of the approach to this same fixed point is not and
depends on the same RG invariant, $\mu$. (It governs the
amplitudes but not the exponents of the operators controlling the
approach to the fixed point.)

This observation in the case of the U(1) Thirring model
is one of the main points of Ref. \cite{tsvelik}.  
It is further pointed out in this work that the dependence
upon $\mu$ can nevertheless be different for different parameter regimes.
In the cases of the $U(1)$ Thirring model, it is essentially
polynomial in $\mu$ for the region termed
AF ($g_{\parallel} < 0$, $g_\perp > 0$, 
$|g_{\parallel}| < g_\perp < \pi - |g_{\parallel}|$), 
but exponential ($\exp(-cst/\mu)$) in the region termed C 
($\pi/2 > g_\perp > 0$, $g_\perp > |g_\parallel |$).
In the latter case therefore,
the dependence on $\mu$ is weak, and so symmetry is, 
in practice, 
certainly restored.
The authors of \cite{tsvelik} then carry on
to conclude that the one loop RG is not reliable, 
as symmetry restoration does sometimes in fact occur 
(in region C), while sometimes it does not (region AF). 

It is here we come to the crux of this work.  As in \cite{tsvelik},
we  consider an alternative 
definition of symmetry
restoration to include all situations where the low energy behaviour
of the theory has only a weak dependence upon the bare anisotropy.
To be clear, we now have two definitions of symmetry restoration
in play:
\vskip .35in 
\noindent{\bf 1. Symmetry restoration induced through 
the renormalization group}.
\vskip .25in
\noindent{\bf 2. Symmetry restoration meaning a weak dependence
of physical quantities on the bare anisotropy.}
\vskip .35in 
\noindent We consider 
now the consequences of this second, expanded definition.
In particular we consider the consequences of combining this definition
with insisting upon a 
field theoretic description of the system. (For the  systems
discussed in this paper, these turn out to be {\it relativistic}
field theories.)

Field theoretic descriptions of condensed matter systems
are desirable both in that
they provide a powerful set of tools and techniques by which
the relevant physics can be extracted, and because they represent
the physics that is universal in nature, i.e. that carries no dependence
upon the microscopic details present in the actual system.
In general, a field theoretical description requires  the parameters
of the model to be in a regime where the
correlation length is much larger than  UV cut-off (the
`lattice spacing'). 
This regime is known as the {\it scaling limit}.
The anisotropic models
considered in this paper are such that they can be studied
exactly in a range of parameters including, but by far exceeding,
the regime where the theory is in the scaling limit.

We will see for the systems considered in this paper
that in order to achieve the scaling limit the bare
parameters must be such that the effects of  
the anisotropy
are small.  The considerations leading to this conclusion
can be phrased in general terms as follows.
We begin with some (possibly free)
theory governed 
by an underlying continuous  symmetry, some
simple  Lie-group, $G$.
Denote the corresponding Lagrangian, $\cal L_G$.
We will then consider perturbations to $\cal L_G$ that, in general,
break the symmetry, $G$, or in an alternative language, anisotropically
deform $G$.  The perturbations will typically take the form,
\begin{equation}
{\cal L}_{\rm pert} = \sum^n_{i = 1} \lambda_i {\cal O}_i ,
\end{equation}
where $n$ is the number of generators of the group, $G$, and typically
${\cal O}_i$ is an operator associated with the i-th generator.
If $\lambda_i = \lambda_j$ for every pair $(i,j)$, the symmetry $G$ is
preserved, while for differing $\lambda_i$, $G$ is broken.  
The initial examples mooted in this paper (interacting Hubbard ladders,
anisotropic Kondo model, and the U(1) Thirring model) all take
this form.

The question of when the theory appears relativistic
is then rephrased:  for what  values of the couplings, $\lambda_i$,
does the theory have a relativistic low energy sector?  
As we will see, this 
requires for the cases considered here
a subset of the bare couplings, $\{\lambda_j\}$, to be taken to
$0$, while the disjoint subset, $\{\lambda_k\}$, is
permitted to be finite.  If it did remain finite, we would
 have the rather odd situation that
\begin{equation}
{\lambda_k \over \lambda_j} = \infty ,
\end{equation}
i.e. the theory would have  an ``infinite bare anisotropy'' 
(defined here by considering the ratios of the bare coupling constants).  
This is unphysical:
no such ratios would be found in a physical system unless enforced
by some symmetry.
But the only possible symmetry, $G$, has by presumption been
broken.  
In order to then remove
this pathology we need to take $\lambda_k \rightarrow 0$ as well. Now,
in the large variety of cases we have considered, it turns out that the 
physical measures of 
the anisotropy in the relativistic limit are determined not by the 
ratios of bare coupling constants, but by quantities which 
are of $O(\lambda)$ or higher.  If all the couplings go to zero
\footnote{It is not however a free theory -- taking this limit is not the
same as setting the couplings to zero.},
it  therefore follows that
the relativistic limit will be isotropic, 
even for arbitrary but  finite ratios of these couplings.  
The lesson we learn is that if 
a model has a relativistic field theoretic description, the latter 
must be isotropic.  Anisotropic relativistic limits will occur only if 
one is willing to accept models with
infinite bare  anisotropy, a rather unphysical requirement.
If we return to our definition of symmetry restoration, we thus
see in taking the scaling limit, symmetry is restored perfectly, 
i.e. there is no dependence on any bare anisotropy.

This, of course, is true only when we apply the scaling limit in the
strict sense.  When we required
$\{\lambda_j\}$ to be zero, we did so to ensure the theory 
looked relativistic
at all possible energy scales.  However this is needlessly restrictive.  We
are only interested in the theory at low energy scales and so are only
interested in it looking relativistic at these same scales.  For example,
in the Kondo model, all we would want is the theory to appear relativistic
on scales less than some large multiple of the Kondo temperature, $T_k$.
Or in the Thirring model, we would only want the model to appear 
relativistic for scales less than some multiple of the fermion mass.  With
such a revised criterion, we find that instead of $\{\lambda_j\}$ being
$0$, they need only be small and finite.  Consequently, the 
set of couplings, $\{\lambda_k\}$, only need to be 
made small and finite in order for the 
ratios,
${\lambda_k \over \lambda_j}$,
to take on reasonable (i.e. physical) values.  This then modifies the 
previous conclusion.  Models that appear anisotropic and 
relativistic at low energies can exist even with finite bare anisotropy,
provided all couplings are
small. 
However, the anisotropy relevant for determining physical
properties
is then extremely small. 
This will be made clear in the examples that are 
found in 
Section II.

The need to make the couplings all small in order to realize
a physical field theoretic description greatly restricts the
amount the low energy behaviour of the theory can deviate from
its isotropic limit.  Thus in taking the scaling limit
we have achieved symmetry restoration under its expanded 
definition (definition 2).
In the cases of U(1) Thirring and anisotropic Kondo
(considered in detail in the
next section together with several other examples)
we know
that the RG invariant
$\mu^2 = g^2_\parallel - g^2_\perp$ governs 
the effect of the anisotropy upon physical quantities
(at least at  sufficiently small energies).
As $g_\parallel$ and $g_\perp$ are
both required to be small
by the scaling limit, $\mu$
is small and so the possible anisotropy is correspondingly small.
For U(1) Thirring, we thus find
that symmetry restoration occurs in both the AF sector (unlike \cite{tsvelik})
and the C sector of the theory.  

Although the two definitions
1 and 2 of symmetry restoration look distinct,
they do share some commonalities.
In both cases there is a concern for the low
energy behaviour of the theory.  And in certain circumstances, the
taking of the scaling limit can be thought of as a crude running of the RG
backward.  In both the AF sector of U(1) Thirring and anisotropic
Kondo, running the RG backwards amounts to taking $g_\perp \to 0$.
This is precisely what the scaling limit requires in both these cases
\footnote{However running the RG backwards does not always mimic
the results of the scaling limit.  In sector C of U(1) Thirring,
the backwards RG flows to a UV fixed point 
different from the point to which the `flow'
of the scaling limit takes one (see the next section).}.
This is not as paradoxical as it might at first seem
(indeed, a reverse RG seems to imply a focus upon the UV not the
IR degrees of freedom in the theory).
In running the RG backwards, i.e. increasing rather than decreasing
the UV cutoff, one removes any distortions the UV cutoff
creates in the low energy sector of the theory.  It precisely
such distortions that render this sector non-relativistic.

We again see a certain complementarity between the two definitions
of symmetry restoration when we understand that
the scaling limit implies a more favourable scenario
for symmetry restoration as understood strictly as an RG induced phenomena.
From the above argument, we know that to even write down a field theoretic
description, we require the couplings to all be small (all, so as to
avoid an unphysically large anisotropy).  Thus the bare theory is only 
weakly anisotropic.  As such, we expect, 
with certain  {\it caveats},
an 1-loop RG predicting 
an enhanced symmetry to be trustworthy: the theory, because it is already
close to being isotropic, will flow onto the manifold of higher
symmetry while the 1-loop description is still valid.

We have already seen one of these caveats in 
operation in the case of $U(1)$ Thirring.
Here the fact that physical quantities depend upon the RG invariant, $\mu$,
means the indicated symmetry restoration does not actually occur.
We, however, conjecture this is something particular to $U(1)$ Thirring.
The U(1) Thirring model possesses a q-deformed quantum
group symmetry, $\widehat{sl(2)}_q$ \cite{BernardLeclair}.  
The parameter, $q$, describing the
symmetry is a function of the parameter, $\mu$, and so
does not change under the
RG.  We thus conjecture that the lack of symmetry restoration is
a reflection of the presence of the $\widehat{sl(2)}_q$ symmetry. 
We will argue (Section IV), however, that if such a
$\widehat{sl(2)}_q$ is explicitly broken
through breaking additionally the $U(1)$ symmetry,
an increase in the symmetry of the problem does occur.
\footnote{We can argue much the same for the low
energy sector of anisotropic Kondo.  However because 
the IR fixed point
is massless, with the anisotropy being {\it irrelevant},
this example is not as compelling.}

A second caveat appears when the RG flow is governed by an
anisotropic (IR) fixed point.
Indeed, here the two definitions of symmetry restoration
do differ in the role assigned to the fixed point of the theory.

The sine qua non of symmetry restoration as an RG-induced phenomena
lies in the nature of the fixed point.
If we have an anisotropic IR  fixed point, we have no expectation
that symmetry restoration according to definition 1 will occur.  
However it may well occur in certain special cases
using our expanded definition 2 of symmetry restoration.
This occurs for instance  in the context of the deformed
O(3) sigma model (the sausage model)
with topological term $\theta = \pi$, which has a line
of fixed points.  Here the RG equations
promise symmetry restoration.  However we know that the particular
point on the fixed line to which the theory flows is determined by the amount
of anisotropy in the theory.  Therefore, symmetry restoration according
to definition 1 does not occur.  However physical quantities depend
only weakly (continuously) on the anisotropy, at least when it is
small,  so that symmetry is restored
according to definition 2.  A similar situation occurs in the
spin-exchange anisotropic $s=1$, 1-channel Kondo model, as discussed in the
concluding section.

If the fixed point is isotropic however,
we allow for the possibility of symmetry restoration under the RG.  
This possibility is realized if the isotropic fixed point
is massless as it is then non-trivial and so determines a set
of physical quantities.  This is the situation
found in the anisotropic $s=1/2$, 1-channel Kondo model.
However if the fixed point
is massive and so trivial, we cannot say with certainty if 
symmetry restoration
under the RG is realized.  With massive flows, it is the approach
to the fixed point that matters, controlled by the nature of
the massive excitations, and it is not possible in general
to characterize this approach.  We do know through the example of
U(1) Thirring that a massive symmetric fixed point does not guarantee
symmetry restoration under the RG.

However if we define symmetry restoration merely as a weak dependence
upon the bare couplings in the low energy sector, the fixed point
need not play an important role.  In both the U(1) Thirring 
and the anisotropic Kondo model we have symmetry restoration so defined 
regardless of
the nature of the fixed point.  Indeed we need not even have an
isotropic fixed point (massless or massive) in order to have
symmetry restoration.  One can easily imagine a scenario in
which the fixed point is anisotropic but the scaling limit restricts
the anisotropy to be weak with a consequent weak dependence of any physical
quantity upon the anisotropy.

The outline of the paper is as follows.  In 
Section II,
we consider
a series of examples illustrating how the scaling limit induces
symmetry restoration (in our expanded sense).
These include the anisotropic Kondo model,
the U(1) Thirring model, a multi-species version of the U(1) Thirring
model, the anisotropic principal chiral model, and
the deformed O(3) sigma model with
$\th = 0$.  All of these
models possess isotropic fixed points.  To explore whether symmetry
restoration occurs when the fixed point is anisotropic, we also consider
in the concluding section
the spin $s>1/2$ anisotropic, 1-channel Kondo model
together with the
deformed O(3) sigma model with $\th = \pi$.

As we indicated, the scaling limit, in providing weak, bare
couplings, promises a more favourable environment for RG-induced
symmetry restoration.
To exploit this conclusion, we consider in 
Section III
a set of theories based upon all possible simple groups 
(the analysis does not extend to semi-simple groups).  In doing
this we extend the results of Lin et al. \cite{fisher}.
We will show that the 1-loop beta functions for such theories all imply an
enhancement in the symmetry.  Given the constraints the scaling limit
places upon the bare couplings, we argue that this enhancement should
be realized in physical models with small enough bare couplings
to admit a field theoretic description. 

Finally in 
Section IV,
we return to the model that lies at the
heart of much of the discussion surrounding the reliability of 
1-loop RG equations, the U(1) Thirring model.
Specifically we
discuss the possibility that
symmetry restoration in the RG sense might take place in
non-integrable variants of the U(1) Thirring model.  We have
conjectured the U(1) Thirring model does not experience symmetry
restoration under the RG flow because of the presence of a
$\widehat{sl(2)_q}$ symmetry.  If this symmetry is then
explicitly broken, an RG induced restoration should be
possible.  And indeed we find it is, although we demonstrate in
the course of the discussion that the matter is
a delicate one.

\section{Interplay of scaling limit and symmetry restoration}

\subsection{The Kondo model}

We start with the anisotropic Kondo Hamiltonian after bosonization: 
$\phi$ is the right moving 
spin field, the charge field having totally decoupled. After standard 
manipulations (see e.g. \cite{GNT}), the Hamiltonian reads
\begin{eqnarray}
\label{firsteq}
H = H_0 &+ {g_\perp\over 4\pi a}\left[s^+ e^{i\sqrt{8\pi}\phi(0)}
+ s^- e^{-i\sqrt{8\pi}\phi(0)}\right] \cr
& + {g_\parallel\over\sqrt{2\pi}}s^z\partial_x\phi(0).
\end{eqnarray}
Here, $s^\pm$ are spin $1/2$ generators, 
$g_\perp,g_\parallel$ the bare Kondo couplings.  In evaluating the 
propagators,
we have used 
$\langle \phi (x) \phi (y) \rangle = -{1\over 4\pi}\log {x-y\over a}$, and 
thus have adopted condensed matter conventions where the cut-off
is left explicitly in the Hamiltonian and the propagator. 
The bulk Hamiltonian is $H_0=v_F \int dx \left[\partial\phi(x)\right]^2$. 

The quickest way
to analyze (\ref{firsteq}) is to perform a canonical transformation with  
$U=\exp\left[i {g_\parallel 
\over v_F \sqrt{2\pi}} s^z \phi(0)\right]$ in order to 
eliminate the $s_z$ term.  The resulting Hamiltonian then reads
\begin{equation}
\label{secondeq}
H = H_0 +
{g_\perp\over 4\pi a}\left[s^+ e^{i\beta\phi(0)}+s^- 
e^{-i\beta\phi(0)}\right],
\end{equation}
where 
$\beta=\sqrt{8\pi}-{g_\parallel\over v_F\sqrt{2\pi}}$. 
In the following, we set $v_F=1$.

Let us now consider $g_\parallel$ in (\ref{secondeq}) as a parameter, 
and set ${\beta^2\over 8\pi} = x < 1$. Here, 
$x$ is the scaling dimension of the exponential operators 
in (\ref{secondeq}). The problem described
by (\ref{secondeq}) will exhibit screening, and by dimensional analysis, 
the Kondo temperature (the temperature that sets the scale by which
the model crosses over from high temperature to low temperature 
behaviour) is 
$T_K\propto {1\over a}\left(g_\perp\right)^{1\over 1-x}$ where again
$a$ is the UV cutoff.  To make the spectrum of this theory purely
relativistic we need to take the field theory - also called  scaling - limit,
$a\to 0$.
At the same time, since  we need to keep an observably finite 
$T_K$, it is  necessary to also take $g_\perp\to 0$.
\footnote{Strictly speaking we are not interested in taking $a \to 0$ so
much as making the ratio, $T_k/(1/a)$ small, that is keeping the 
energy scale in the problem governing the physics small compared to
the bandwidth.  Only in doing so do we expect to find a relativistic
theory free of distortions from the bandwidth of the theory.  However
one achieves the same effect by taking $a \to 0$ while insisting
the energy scale $T_K$ remains finite.}

It is useful to stress here that the canonical transformation used in going 
from (\ref{firsteq}) to (\ref{secondeq}) is valid in the scaling limit only. 
In this limit,
the somewhat different aspects of the $SU(2)$ Kondo Hamiltonian are readily 
explained by the foregoing points: in (\ref{firsteq}),
the isotropic theory has $g_\parallel=g_\perp$, while in (\ref{secondeq}), 
it is
described by  $\beta=\sqrt{8\pi}$, that is, $g_\parallel=0$. 
However, as one requires $g_\perp\to 0$,
these two points of view are completely equivalent.

In the scaling limit, the Kondo problem becomes describable
by an integrable massless field theory. 
The bulk excitations have dispersion relations
$p=e=M e^\theta$, with $\theta$ the rapidity,
$M$ an energy like parameter which has no physical significance since the 
bulk theory is massless.  They
have a factorized scattering described by a solution of the Yang-Baxter and 
bootstrap equations.  Among these 
excitations, the most important are the kink and antikink.  Physical 
properties depend only on the ratio $T/T_K$, where $T$ is the physical 
temperature.
The kinks/antikinks scatter off the Kondo impurity with an amplitude 
described by $R=-i\tanh\left({\theta-\theta_K\over 2}-{i\pi\over 4}\right)$,
where $T_K=M e^{\theta_K}$ (here we see that changes in the mass scale $M$
can be readily absorbed in a shift of the rapidities).  They also scatter 
among one another,
their scattering being determined by 
a six vertex like solution of the Yang-Baxter equation.
This solution, in turn, has a quantum affine algebra 
symmetry $\widehat{sl(2)}_q$
(we follow here the conventions of \cite{BernardLeclair}, for which
$q=-1$ is the isotropic limit),
where in this case $q=e^{-{i\pi\over x}}$. 
For small values of $g_\parallel$, $q\approx -e^{i g_\parallel/2}$

To summarize, the Kondo 
anisotropic field theory is  obtained as $g_\perp\to 0$, and it is 
fully  characterized by the value of $x<1$, 
that is the value of $g_\parallel$.
We thus come to the conclusion as stated in the introduction:

\smallskip
\bigskip

\noindent
{\it To observe a finite anisotropy in the field theory ($x\neq 1$), 
one needs to have an infinite anisotropy 
in the bare theory ${\cal A}={g_\parallel/ g_\perp}\to\infty$ as $a\to 0$. }

\smallskip
\bigskip

\noindent We also note that if we maintain a finite anisotropy in taking
the scaling limit, that is, 
${\cal A}={g_\parallel / g_\perp} = {\rm constant}$ (and $g_\parallel>0$) 
as $g_\perp \rightarrow 0$
while $a\to 0$,
the model will be described by the same isotropic
$SU(2)$ Kondo field (scattering) theory.

The foregoing discussion avoided the issue of regularization, 
about which we would like
to comment now.  
When $g_\parallel$ is large enough, 
i.e. $\beta< \sqrt{4\pi}$, no further
renormalization is needed besides the usual normal ordering for 
vertex operators (implicit in 
all our notations). 
When $\sqrt{4\pi}< \beta<\sqrt{8\pi}$, an extra subtraction is necessary,
since divergences appear in the computation of the impurity free 
energy (in complete analogy
with the bulk sine-Gordon model). Our conclusions then apply to the 
universal part of the
physical properties of interest. 

Regularization issues are most cleanly controlled in the framework
of the exact solution of the Kondo model \cite{wiegmann}, for which we now
would like to recast our arguments. 
The Hamiltonian is as in (\ref{KH}).
In order to do so we first need to consider the excitations
in this theory.  It is well understood that the lowest energy
spin and charge excitations in the model are decoupled with the
charge excitations completely trivial (as is to be expected
given the bosonization of the problem).  We thus focus on the spin
excitations.  At $T=0$, 
their energy, $e(\lambda )$, and momentum, $p(\lambda )$,
\footnote{Here 
$\lambda$ is a rapidity and is
not to be confused with the coupling constants 
discussed in the introduction.}
are given by 
\begin{equation}
e(\lambda ) = -\int_Q d\lambda' e(\lambda')
\del_{\lambda'}\theta_2(\lambda'-\lambda )- {N\over L}
(\theta_1 (\lambda) + \pi),
\end{equation}
and 
\begin{equation}
p(\lambda ) = -\int_Q d\lambda' p(\lambda')
\del_{\lambda'}\theta_2(\lambda'-\lambda )
+ {N\over L}\theta_1 (\lambda),
\end{equation}
where
\begin{equation}
\theta_n(\lambda ) = 2\tan ^{-1} [\tanh (\mu \lambda )\cot (n\mu/2)].
\end{equation}
N is the number of particles in the system, and $L$ is the length
of the system.  Here $Q$ represents the interval of the spectral
parameter, $\lambda$, (or, more intuitively, the spin rapidity)
over which (spin) excitations are present in the ground state.
The parameter, $\mu$, together with another parameter, $f$, serve to
characterize the anisotropy in the system.  As functions of $g_\perp$
and $g_\parallel$ they are given by
\begin{equation}\label{barerel}
(\coth f)^2={\sin^2 \left({g_\parallel\over 2}\right)\over 
\sin\left({g_\parallel+g_\perp\over 2}\right)
\sin\left({g_\parallel-g_\perp\over 2}\right)},
\end{equation}
and
\begin{equation}\label{otherbarerel}
\cos\mu={\cos(g_\parallel/ 2)\over \cos(g_\perp/ 2)}.
\end{equation}
Isotropy is reached by taking $\mu , f \rightarrow 0$ with 
$f/\mu = {\rm constant}$.  We restrict ourselves to the region
where $\pi - g_\perp > g_\parallel > g_\perp > 0$ such that 
$\mu$ is a real parameter.
Finally, we can express the Kondo temperature in terms of
the anisotropy:
\begin{equation}
T_K = {2 E_F \over \pi} e^{-{\pi f\over \mu}},
\end{equation}
where $E_F$ 
is the Fermi energy 
(proportional to the inverse of the UV cutoff) of the theory.

In taking the scaling limit, we need to ensure the low energy excitations
look relativistic.  We see that the above integral equations reduce to
\begin{equation}\label{dispersion}
e(\lambda ) = c - p(\lambda ),
\end{equation}
that is, the energy-momentum already obey a relativistic dispersion
relation.  However we still need to ensure the energy scale in
the system, $T_k$, is far below  the bandwidth.  Thus we
need to take $T_k/E_F \rightarrow 0$ in order to achieve the scaling
limit.
For finite $\mu$ this requires
$f\to\infty$.  In that limit, the foregoing equations simplify
and one finds: 
\begin{equation}
e^{-f}\approx {{g_\perp\over 4} \cot{g_\parallel\over 2}},~~~ \mu\approx 
{g_\parallel\over 2}.
\end{equation}
The same qualitative conclusions therefore hold; that is, to have a 
finite anisotropy in the scaling limit,
one needs to send $g_\perp$ to zero, while $g_\parallel$ remains finite, hence 
giving rise to an
infinite bare anisotropy. One also checks that 
$T_K\propto 
{1\over a}\left(g_\perp\right)^{\pi\over\mu}\propto 
{1\over a}\left(g_\perp\right)^{1\over 1-x}$. 
If instead we have a finite anisotropy, i.e.  let $g_\parallel$ and $g_\perp$ 
both go to zero
while $g_\parallel/g_\perp\to\gamma$, $\gamma$ a finite number, we get 
that $\coth^2 f\to {\gamma^2\over\gamma^2-1}$,
while $\mu\to 0$, corresponding to an isotropic field theory whatever 
the value of $\gamma$. In that limit, one has 
$T_K\propto {1\over a}e^{-{\pi f\over\mu}}
\propto {1\over a}e^{-{cst\over g_\perp}}$, a well known result 
for the isotropic Kondo model. The exact solution 
thus fully confirms the foregoing field theoretic analysis.

Now, as we indicated in the introduction, enforcing the scaling limit
strictly is unnecessary from the condensed matter point of view.  
All we require is that the low energy excitations
in the theory appear relativistic.  Returning then 
to the exact solution of the Kondo model,
instead of $T_K/E_F\to 0$, we require that
\begin{equation}
{T_K \over E_F} \leq {1 \over A}
\end{equation}
where A is some large number.  In terms of the parameters $f$ and $\mu$,
this translates into the condition,
\begin{equation}
{f \over \mu} \geq {1\over \pi} \log {2A\over\pi}.
\end{equation}
With this condition it no longer is necessary to have 
$g_\perp \rightarrow 0$.  Rather, a region of parameter space of finite
area satisfies the above constraint.  We have plotted this region in Figure
1 for $A=50$ (that is, the theory is relativistic for scales up to $50T_K$).

We observe that the region in Figure 1 is restricted to less than one-half
of the possible parameter space.  If we insist that 
$g_\parallel \geq g_\perp \geq g_\parallel/2$ (what we may call a 
maximal `reasonable'
value for the ratio of couplings), the 
region of parameter space is restricted to
the gray shaded region between the two lines.  We thus see that any
physical realization 
of the Kondo model with a scaling limit will not be far from the 
isotropic ray $g_\parallel = g_\perp$ .

Yet the fraction of parameter space is still appreciable.  In part this is
an artifact of the initial data of the analysis of the Kondo model 
\cite{wiegmann}: this
analysis begins with a linear spectrum, and as there are no bulk interactions
in this model, the spectrum remains linear as evinced in (\ref{dispersion}).
It is possible to consider a version of the Kondo model that
is equipped with band structure.  One can do this straightforwardly
by analyzing a lattice system with a Kondo impurity
so that the initial bulk electron spectrum 
obeys a dispersion relation of the form, $\epsilon (k) = -2t \cos (k)$.
Or one can do this more abstractly 
by turning to an integrable lattice regularization of the Kondo 
spin dynamics \cite{HSunpub}. 
The latter follows from a general 
construction, where one introduces a line of spectral parameter 
defects in an otherwise homogeneous 
6-vertex model \cite{HSunpub}.  
The equations that arise from this latter construction
have a form similar to those of \cite{wiegmann}, except for the fact that 
the bare energy, instead of being given by $e\propto \theta_1(\lambda)$,
now has the form $e\propto {d\over d\lambda} \theta_1(\lambda)$. 
The introduction of this curvature shrinks the scaling region
considerably, leading to results similar to those of the $U(1)$ 
Thirring model to be discussed next where the bulk perturbation
of this system also yields a non-linear dispersion relation. 

\begin{figure}[tbh]
\centerline{\psfig{figure=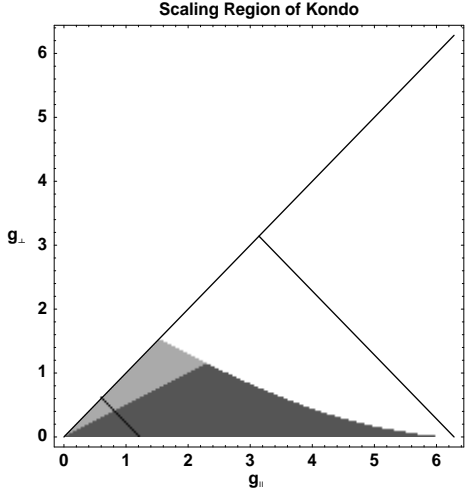,height=4in,width=3in}}
\caption{The portion of parameter space where the Kondo model
is well described by a field theory.  The lower triangle marks out
the total region of considered parameter space (a quarter of the total).  
The shaded region marks the area
where the theory is relativistic while the more lightly shaded subarea 
is characterized by $g_\perp > g_\parallel /2$.}
\end{figure}

Given the available parameter space in Figure 1, 
we can now ask how the physics of the problem
varies as we vary the anisotropy.  With this in mind we consider the impurity
susceptibility, $\chi (H)$, as a function of magnetic field.  From
\cite{wiegmann}, we know it is given by
\begin{eqnarray}
\chi(H<T_H) &=&
{1\over H \sqrt{\pi}}\sum^\infty_{n=0} {(-1)^n\over n!}
e^{(2n+1)\pi a(\mu )}\times \cr
&& ({H\over T_H})^{2n+1} {\Gamma (1+{\pi\over\mu}(n+{1\over2})) \over
\Gamma (1+({\pi\over\mu}-1)(n+{1\over2}))};\cr
\chi(H>T_H)&=&{\mu\over\pi^{5/2}}\sum^{\infty}_{n=1} 
{1\over (n-1)!} \sin (\mu n)\times\cr
&& \hskip -.6in \Gamma ({1\over 2} + {\mu n \over \pi})
\Gamma (n(1 - {\mu\over\pi}))
e^{-2\mu na(\mu )} ({H\over T_H})^{2\mu n\over\pi}, 
\end{eqnarray}
where 
\begin{equation}
a(\mu ) = {1\over 2\mu}\log\left(1-{\mu\over\pi}\right) - 
{1\over 2\pi}\log \left({\pi\over\mu} -1\right),
\end{equation}
and $T_H$ is related to the Kondo temperature, $T_K$, via
\begin{equation}
T_H = 2\sqrt{\pi}{\Gamma (1+{\pi\over 2\mu})\over \Gamma (1/2+{\pi\over 2\mu})}
e^{\pi a(\mu )} T_K.
\end{equation}

\begin{figure}[tbh]
\centerline{\hskip -.5in
\psfig{figure=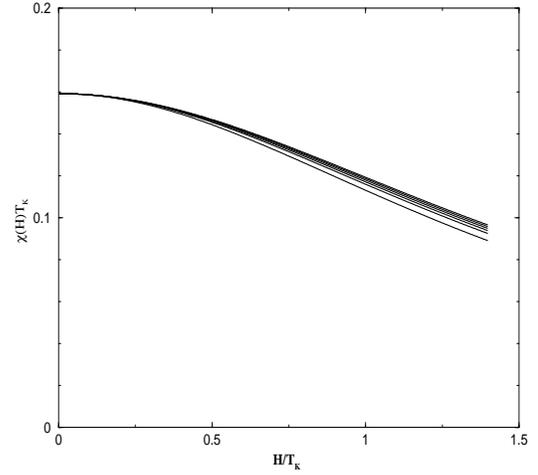,angle=-90,height=2.5in,width=2.5in}}
\caption{Behaviour of the magnetic susceptibility as a function
of the anisotropy for small magnetic field, i.e. $H < T_H$.  From
bottom to top, the plots correspond to 
${5 \over \pi} (g_\parallel , g_\perp) = (1,1), (6/5,4/5), (7/5,3/5),
(8/5,2/5), (9/5,1/5),$ ${\rm and}~(2,0)$.}
\end{figure}

For small fields, $\chi (H) \sim T_K^{-1}$.  We thus scale out this factor
and plot $\chi (H)T_K$ as a function of $H/T_K$ in Figures 2 and 3 for
different pairs of $(g_\parallel,g_\perp)$.  
The points chosen fall equidistantly on a line connecting
$g_\parallel = g_\perp = \pi/5$ with
$g_\parallel = 2\pi/5 , g_\perp = 0$.  (This line is marked
on Figure 1).  We see that the appropriately
scaled $\chi$ varies only slightly when the isotropy of the model
is deformed, even if the deformation is strong.

\begin{figure}[tbh]
\centerline{\hskip -.5in
\psfig{figure=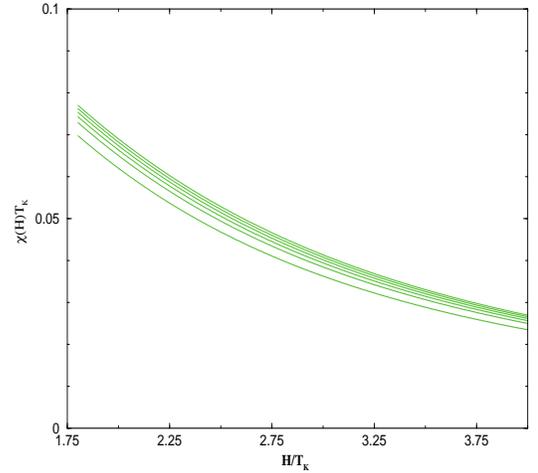,angle=-90,height=2.5in,width=2.5in}}
\caption{Behaviour of the magnetic susceptibility as a function
of the anisotropy for large magnetic field, i.e. $H > T_H$.
From bottom to top, the plots correspond to 
${5 \over \pi} (g_\parallel , g_\perp) = (1,1), (6/5,4/5), (7/5,3/5),
(8/5,2/5), (9/5,1/5),$ ${\rm and}~(2,0)$.}
\end{figure}

\noindent It is unsurprising that the low field 
susceptibility in Figure 2 varies only a little 
given the anisotropic Kondo model shares the same fixed point as its 
isotropic counterpart.  One
can see the curves collapse upon one another as $H\rightarrow 0$ indicative
of the flow to this same IR fixed point.  However the variation is
also small for the high field case in Figure 3 where one ostensibly expects
to be far from the fixed point.  

To conclude, we see that the scaling limit has drastic consequences
for how isotropic a theory is.
In order for a system with given bare coupling 
constants and finite
bare anisotropy,
to be reasonably described by the Kondo field theory,
it must be only weakly anisotropic.
The variation of the physical quantities 
over the allowed anisotropy is correspondingly small.
Excluding the
possibility of extraordinarily (i.e. unphysically) 
strong anisotropy in the bare coupling 
constants, the scaling limit thus enforces a strong restoration of symmetry
in its expanded sense.

\subsection{The U(1) Thirring Model}

We now move on to consider the U(1) Thirring model where an analogous series
of conclusions to those of the Kondo model will be drawn.  
In fact, this model is essentially a bulk version of the anisotropic 
Kondo model of the previous section.

\begin{figure}[tbh]
\centerline{
\hskip .4in\psfig{figure=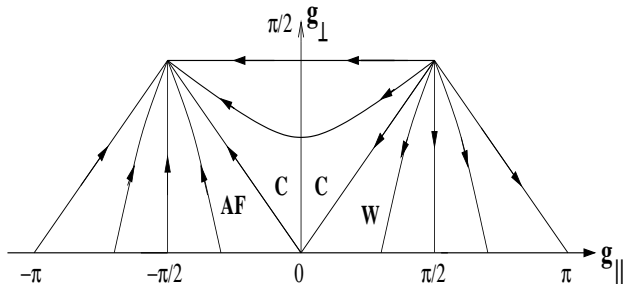,height=1.5in,width=3.5in}}
\caption{The phase diagram of the U(1) Thirring model.  Marked are three
regions of primary interest: AF, C, and W.}
\end{figure}

The U(1) Thirring model is described by the Lagrangian 
\begin{eqnarray}
\cal{L} &=& i\bar\psi_\alpha\gamma_u\del^\mu\psi_\alpha
+ {1\over 4}g_\parallel (j_z)^2 \cr
&&\hskip .5in + {1\over 4}g_\perp 
\left[(j_x)^2+(j_y)^2\right] ,
\end{eqnarray}
where $j^a_\mu = \bar\psi_\alpha\gamma_\mu\tau^a_{\alpha\beta}\psi_\beta$
and $\psi$ is a doublet of Dirac spinors.  This model was completely 
solved by algebraic Bethe ansatz \cite{jap}.  

There, in accordance with earlier perturbative
work, it was established that according to the values of the couplings
$g_\perp$ and $g_\parallel$, the model exhibits 
different infrared behaviour.
The phase diagram, pictured in Figure 4, divides into three sectors of
interest.  In
sectors AF (asymptotic freedom) and C (crossover), a strong coupling regime
appears and there is dynamical mass generation.  The flows in both
these sectors are towards a stable IR fixed point 
(located at $(g_\parallel, g_\perp)=(-\pi/2, \pi/2)$ in Fig. 4).

The two sectors AF and C
are distinguished by their behaviour in the UV: 
in the AF region the theory is asymptotically free while the UV flow
in region C is to a strongly coupled fixed point.
In the AF sector (at weak coupling) the model maps onto the sine-Gordon
theory.  The sine-Gordon theory can be thought of as an anisotropic
current-current perturbation of a free boson theory.  Thus the U(1)
Thirring model in the AF sector bares some resemblance to the anisotropic
Kondo model, itself representable as a free boson perturbed 
by an anisotropic current-spin interaction.

In the final sector, W, the weak coupling sector, the $jj$ 
perturbation is irrelevant and the theory can be described perturbatively.
In this sector, the excitations are all massless.  As we are ultimately
interested in looking at theories with flows implying symmetry
restoration, we will focus on regions AF and C.

\subsubsection{AF Region}

We start with the AF region.
As with the Kondo model, the excitations divide into decoupled spin and
charge sectors, with the charge sector trivial.  The spin excitations, at
least in the regime that will be of interest, are given from the
following integral equations:
\begin{eqnarray}\label{u1eq}
e(\lambda ) &+& \int^B_{-B}R(\lambda -\lambda')e(\lambda ')d\lambda'
\!=\! {H \over 2(1\!-\!{\mu\over\pi})} - e_0(\lambda ) ,\cr
p(\lambda ) &+& \int^B_{-B}R(\lambda -\lambda')p(\lambda ')d\lambda'
\!=\!  - p_0(\lambda ),
\end{eqnarray}
where $e_0(\lambda )$ and $p_0(\lambda )$ are
\begin{eqnarray}
e_0(\lambda ) &= {\Lambda_c \over 2}\tan^{-1} 
\left[{\cosh ({\pi\lambda \over 2}) \over 
\sinh ({\pi f\over 2\mu})}\right];\cr
p_0(\lambda ) &= {\Lambda_c \over 2}\tan^{-1} 
\left[{\sinh ({\pi\lambda \over 2}) \over 
\cosh ({\pi f\over 2\mu})}\right].
\end{eqnarray}
Here $\Lambda_c$ is proportional to the bandwidth and $f$ and $\mu$
are related to the bare parameters of the model, $g_\perp$ and 
$g_\parallel$,
in a near identical fashion to the relations (\ref{barerel}) and 
(\ref{otherbarerel}):
\begin{eqnarray}\label{u1rel}
\coth^2 (f) &=& {\sin^2 (g_\parallel) \over \sin (g_\parallel+g_\perp)
\sin (g_\parallel - g_\perp)}\cr
\cos (\mu ) &=& {\cos (g_\parallel ) \over \cos (g_\perp )}.
\end{eqnarray}
$H$ is the magnetic field in the problem, and $R(\lambda )$, the kernel
of the integral equations, is given by
\begin{equation}
R(\lambda ) = -{1\over 2\pi} \int^\infty_{-\infty} 
{\sinh( ({\pi\over\mu } -2)\omega) \over 2\cosh (\omega )
\sinh \left[({\pi\over\mu}-1)\omega \right] }.
\end{equation}
The limit, $B$, the Fermi rapidity,
in the integral equation is such that
\begin{equation}
e(\lambda ) \geq 0 ~~~~ {\rm for~|\lambda |\leq B}.
\end{equation}
When $H=0$, $e(\lambda ) \leq 0$ and $B=0$.  In this case the ground state
of the system is filled with states for all $\lambda$.  Elementary
excitations are obtained by creating holes with energy-momentum
$(e_0 (\lambda ), p_0(\lambda ))$ in this filled sea.

We must consider values of the parameters $(|g_\perp| < |g_\parallel|
< \pi - |g_\perp|)$ such that $f$ and $\mu$ are real.  
To make the spectrum relativistic,
we need to again take the scaling limit.
If we hold $\mu$ finite, that is, maintain a finite  anisotropy
even in the scaling limit, we require $f \rightarrow \infty$.  This, 
as before,
requires $g_\perp \rightarrow 0$, and thus a ratio of bare coupling 
constants  $g_\parallel / g_\perp \to \infty$, i.e. 
an infinite bare anisotropy. 
Meanwhile, if we keep a finite bare anisotropy, 
both bare couplings have to go to zero 
in the scaling limit, in which case $f$ remains finite 
while $\mu \to 0$, i.e. the field theory is 
isotropic.  These conclusions are exactly the same as the ones 
obtained in the Kondo case. 

\begin{figure}[tbh]
\centerline{\psfig{figure=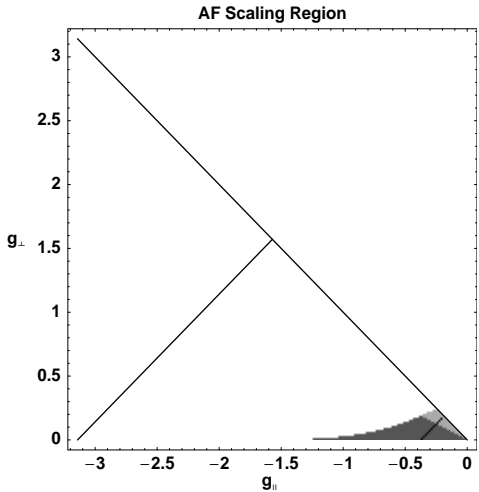,height=4.in,width=3in}}
\caption{Region in the (AF) portion of 
parameter space where the U(1) Thirring model
is well described by a field theory.  The lower triangle marks the
total region of parameter space considered while the shaded region
marks the portion where the theory is relativistic.  The more lightly
shaded subregion is given by the condition that $g_\perp > |g_\parallel|/2$.}
\end{figure}
 
However as before, we do not need to take a strict scaling limit.
With $f/\mu \gg 1$ and $H=0$, the above dispersion relation becomes
\begin{eqnarray}\label{disu1af}
e(\lambda ) &= m \cosh ({\pi\lambda \over 2})(1 + {\rm correction~terms});\cr
p(\lambda ) &= m \sinh ({\pi\lambda \over 2})(1 + {\rm correction~terms}),
\end{eqnarray}
where $m = \Lambda_c e^{-\pi f /{2\mu}}$.  
In relaxing the scaling limit, we require
instead that
\begin{equation}
{\rm correction~terms} < {1\over A},
\end{equation}
for all $\lambda$ such that $e(\lambda ) < Cm$, that is the
spectrum appears relativistic to one part in A up to scales of $C\times m$.
Imposing the constraints leads to the condition
\begin{equation}
{f \over \mu} \geq {1\over\pi} \log ({4C^2A\over 3}).
\end{equation}
In Figure 5 the shaded region marks 
the values $(g_\parallel , g_\perp)$ permitted
by the above relationship for $A=C=50$.  
Unlike Figure 1 describing the allowed parameter
space for the Kondo model, the region in Figure 5 is far smaller.
This is a consequence of the exact expressions for energy-momentum not being
relativistic from the start.  The need to make the correction terms small
leads to the large reduction in allowed parameter space.  Because
of the log dependence in the above constraint, changing A and C does
not drastically affect the allowed region of parameter space.

\begin{figure}[tbh]
\centerline{\hskip -.5in 
\psfig{figure=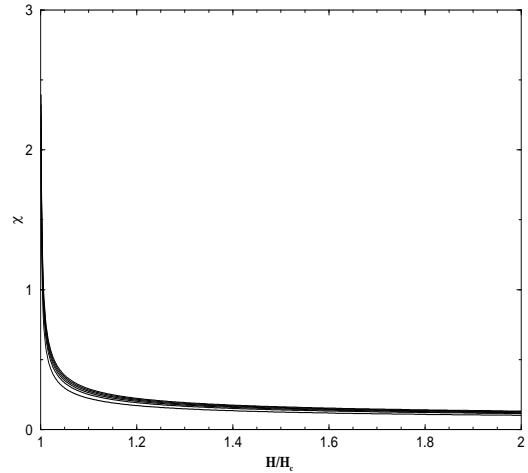,angle=-90,height=2.5in,width=2.5in}}
\caption{The dependence of the susceptibility in the AF region upon
the anisotropy.  From bottom to top, the plots correspond to
six equally spaced values of $(g_\parallel,g_\perp)$ along the
line $(-.2,2.)$ to $(-.4,0)$.}
\end{figure}

Again we ask how the magnetic susceptibility $\chi$ depends upon the
anisotropy.  The magnetic susceptibility in the AF sector can be given in
terms of the energy:
\begin{equation}
\chi(H) = -\del^2_H E(H),
\end{equation}
where $E(H)$ is the ground state
energy per unit length as a function of $H$:
\begin{equation}
E(H) - E(0) = -{\Lambda_c \over 2}\int^B_{-B} d\lambda
e(\lambda ) \rho_0 (\lambda ),
\end{equation}
where $\rho_0$ is the bare density of states whose low energy
behaviour (valid whenever (\ref{disu1af}) holds)
is governed by $\rho_0 = (m/ 2\Lambda_c)\cosh (\pi\lambda /2)$.
For small H, the integral equation for $e(\lambda )$ can be solved
iteratively with the result \cite{jap}
\begin{eqnarray}\label{afchi}
\chi (H) &=& {1\over 2^{5/2} \pi} {1\over (1-\mu/\pi)^2}
\bigg\{ ({H-H_c \over H_c})^{-1/2}\cr 
&& - {2^{9/2} \over \pi}R(0) + {\cal O}({H-H_c \over H_c}) \bigg\},
\end{eqnarray}
where $R$ refers to the kernel in (\ref{u1eq}) and
\begin{equation}
H_c = 2m(1-\mu/\pi ) .
\end{equation}
In Figure 6 we plot $\chi (H)$ for different values of $g_\perp$ and
$g_\parallel$ ranging from $g_\perp = -g_\parallel = .2$
along a straight line to $g_\perp = 0$, $-g_\parallel = 0.4$ (shown
in Figure 5).  We see the
plots as a function of $H/H_c$ fall nearly on top of one another.  Thus in the
region where a field theoretic description is possible, the allowed anisotropy
does not lead to drastic variations in the physics.

\subsubsection{Region C}

We now consider sector C.
For the region $|g_\parallel| < |g_\perp| < \pi /2$,
the same relations as found in (\ref{u1rel}) hold, but here the couplings
$\mu$ and $f$ are purely imaginary.  We thus set $\mu_1 = -i\mu$ and
$f_1 = if$ with the result
\begin{eqnarray}\label{u1crel}
\cot^2 (f_1) &=& {\sin^2 (g_\parallel) \over \sin (g_\parallel+g_\perp)
\sin (g_\parallel - g_\perp)} ,\cr
\cosh (\mu_1 ) &=& {\cos (g_\parallel ) \over \cos (g_\perp )}.
\end{eqnarray}
In this sector the spin excitations are described by analogous integral
equations as those for the AF sector (\ref{u1eq}), but with $e_0$ and $p_0$
replaced by
\begin{eqnarray}
e_0(\lambda ) &= {\Lambda_c \over 2}\sum^\infty_{l=-\infty} \tan^{-1} 
\left[{\sinh ({\pi f_1\over 2\mu_1}) 
\over 
\cosh \left({\pi\over 2}(\lambda - {2\pi l\over \mu_1})\right)}\right] ,\cr
p_0(\lambda ) &= {\Lambda_c \over 2}\sum^\infty_{l=-\infty} \tan^{-1} 
\left[{\cosh ({\pi f_1\over 2\mu_1}) 
\over \sinh \left({\pi\over 2}(\lambda - {2\pi l\over \mu_1})\right)}
\right] ,
\end{eqnarray}
and $R$, the kernel, now given by
\begin{eqnarray}\label{u1cr}
R(\lambda ) &=& 
\sum^\infty_{n=-\infty} {\tilde R}(\lambda - {2\pi n\over \mu_1});\cr
{\tilde R}(\lambda ) &=& -{1\over\pi}\int^\infty_0 
d\omega {\cos(\omega\lambda ) \over 1 + e^{2\omega}}.
\end{eqnarray}
Again when $H=0$, the energy-momentum of the excitations is simply
$(e_0(\lambda ),p_0(\lambda ))$.  Note that $\tilde{R} (\lambda )$ is the
isotropic limit of the kernel $R$ in the AF sector.  Thus the
sum forming $R$ here represents the isotropic limit $(n=0)$ plus
what will turn out to be exponentially small corrections $(n\neq 0)$.
Because the structure of these expressions for the energy-momentum
is considerably more complicated, we need to break down the analysis into
two cases: $-\pi/2 < f_1 < 0$ and $0 < f_1 < \pi/2$.  We will always
assume $\mu$ is such that $\exp (-\pi^2/2\mu) \ll 1$.

In the case, $-\pi /2 < f_1 < 0$, the lowest energy excitations
correspond to those for $\lambda \sim 0$.  Expanding about this point,
$e(\lambda )$ becomes
\begin{equation}
e(\lambda ) = -\Lambda_c \pi + m\cosh ({\pi\lambda \over 2}) \times
  (1+{\rm correction~terms}),
\end{equation}
where $m = \Lambda_c\exp (-\pi |f_1|/2\mu_1 )$.  
In this case some of the correction
terms are of ${\cal O}(\exp(-\pi^2/\mu_1 ))$.  Thus in order to take the
true scaling limit here (i.e. take the correction terms to zero), we
need to take $\mu_1 \rightarrow 0$.  In the process, we see from
(\ref{u1crel}) that 
{\bf both} $g_\parallel$ and $g_\perp$ have to go to zero, while their 
ratio can be an arbitrary number.  In this region of coupling constants 
therefore,
the continuum limit, when it exists, is always isotropic. 
This is not so different that the
results encountered in the AF region of parameter space.
In region C the ratio of couplings, $|g_\parallel|\leq |g_\perp|$,
is always finite.  In region AF the finiteness of this ratio leads
to an isotropic scaling limit. 

\begin{figure}[tbh]
\centerline{\psfig{figure=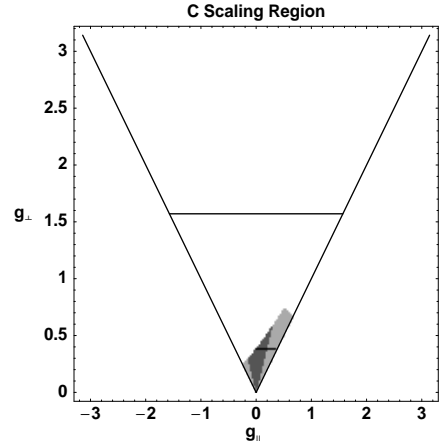,height=4.in,width=3in}}
\caption{The portion of region C where the U(1) Thirring model
is well described by a field theory.  The lower triangle marks out
the total considered parameter region while the shaded region is
where the theory is relativistic.  The more lightly shaded area
meets the additional condition that $g_\parallel > g_\perp/2$.}
\end{figure}

In relaxing the scaling limit to insisting the correction terms are $< 1/A$
for scales up to $C\times m$, we instead find that the 
following two conditions
need to be satisfied:
\begin{eqnarray}
{f_1\over \mu_1} &\geq& {1\over \pi} \log {4C^2 A \over 3};\cr
{(\pi - f_1)\over \mu_1} &\geq& {1\over \pi} \log {A}.
\end{eqnarray}
The region of parameter space satisfying these conditions 
for $A=C=50$ is a portion of the shaded region 
in Figure 7.

In the case $0 < f_1 < \pi /2$, the lowest energy excitations occur
nearest $\lambda \sim \pm\pi/\mu_1$.  Near these points, $e(\lambda )$
takes the form
\begin{equation}
e(\pm{\pi\over\mu}-\lambda ) = m\cosh \left({\pi\lambda \over 2}\right)
(1+{\rm correction~terms}),
\end{equation}
with $m = \Lambda_c\exp ({\pi\over 2\mu_1}(\pi-f_1))$. 
Again the correction
terms are of ${\cal O}(\exp({-\pi^2/\mu_1}))$ and hence 
$\mu_1 \rightarrow 0$ in the scaling limit thus restoring isotropy.
If we relax the scaling limit we find an analogous set of constraints
as above:
\begin{eqnarray}
{(\pi - f_1)\over \mu_1} &\geq& {1\over \pi} \log {4C^2 A \over 3};\cr
{f_1\over \mu_1} &\geq& {1\over \pi} \log {A}.
\end{eqnarray}
The region of parameter space satisfying these constraints 
completes the shaded region in Figure 7.

We see that the total permitted
region is asymmetric under $g_\parallel \rightarrow -g_\parallel$.
This underlying asymmetry is reflected in the RG: the RG flows
from an UV fixed point at negative $g_\parallel$ to an IR fixed point
at positive 
$g_\parallel$.  The scaling region is weighted towards a parameter
regime away from the UV fixed point as it is in this regime
that the low energy sector of the theory is least distorted by
the UV cutoff.
Unlike the AF sector, a good part of the region meeting these
constraints is such that $g_\perp /g_\parallel < 2$ (the region shaded
gray in Figure 7), that is, the couplings here take on reasonable values.

\begin{figure}[tbh]
\centerline{\hskip -.5in
\psfig{figure=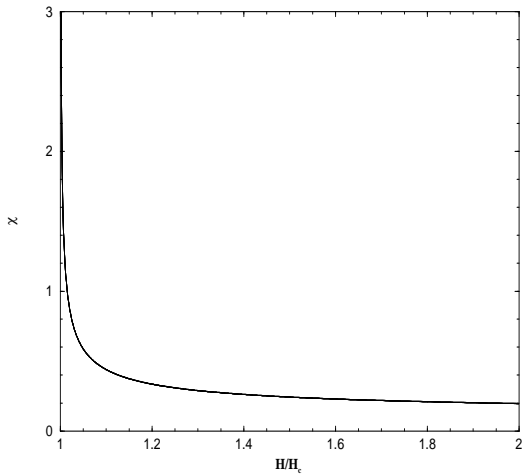,angle=-90,height=2.5in,width=2.5in}}
\caption{The dependence of the susceptibility in the C region upon
the anisotropy.  The plots correspond to
six equally spaced values of $(g_\parallel,g_\perp)$ along the
line $(.4,.4)$ to $(0,.4)$.}
\end{figure}

We again ask how the magnetic susceptibility varies over the permitted
anisotropy.  The susceptibility takes on a similar form to that in
(\ref{afchi}),
\begin{eqnarray}\label{cchi}
\chi (H) &=& {1\over 2^{5/2} \pi}
\bigg\{ ({H-H_c \over H_c})^{-1/2}\cr 
&& - {2^{9/2} \over \pi}R(0) + {\cal O}({H-H_c \over H_c}) \bigg\},
\end{eqnarray}
but with $H_c = 2m$ (its isotropic value) and $R$ the kernel
in (\ref{u1cr}).  We note that the first term in the susceptibility has
no dependence upon $\mu$.  To leading order, region C is thus identical
to the isotropic model.
In Figure 8 we plot how $\chi (H)$ varies when the
couplings are changed for six equally space points 
along the line from ($g_\perp = .4 , g_\parallel = .4$)
to ($g_\perp = .4 , g_\parallel = 0$) (plotted in Figure 7).  
We see there is no discernible
variation in $\chi$ from its isotropic value.  This is a consequence of
the kernel in (\ref{u1cr}).  Terms for $n\neq 0$ correspond to
exponentially small corrections to the isotropic limit.

\subsection{MultiFlavour Fermion Model}

In this section we consider a multiflavour fermion variant of the 
U(1) Thirring model
and show that the same conclusions hold.
In particular, 
we demonstrate that the symmetry restoration is not affected
by the number of fermion species in the theory.
 
The action, involving $2s+1$ fermion
species, that we consider is,
\begin{eqnarray}\label{multiS}
S &=& \int d^2x \big(\sum^{2s+1}_{a=1} i \bar\psi_a\gamma_\mu\del^\mu\psi_a
+ \cr
&& \hskip 1in\bar\psi_{a_2}\gamma_\mu\psi^{a_1}V^{a_2b_2}_{a_1b_1}
\bar\psi_{b_2}\gamma^\mu\psi^{b_1}\big).
\end{eqnarray}
The interaction
potential, $V$, is chosen such that the model is integrable.  For $s > 1/2$,
V has a complicated functional form.  But in the scaling limit, the
interaction takes on the simple form,
\begin{equation}
V = J_\parallel S^z_1 S^z_2 + J_\perp S^1_\perp S^2_\perp,
\end{equation}
where $S^a$ are the generators of the spin-$s$ representation
of $SU(2)$.

The fermionic model in (\ref{multiS}) has been solved with Bethe
ansatz \cite{babu}.  The structure of the solution mimics that of
U(1) Thirring in the AF region.  Thus there are spin and charge
excitations with the charge excitations trivial.  The lowest
energy excitations are given by
\begin{eqnarray}
e(\la ) + \int^B_{-B} d\la' R(\la -\la') e(\la') &=& {H \over 2(1-2s\mu /\pi )}
- e_0 (\la );\nonumber\\
p(\la ) + \int^B_{-B} d\la' R(\la -\la') p(\la') &=& - p_0 (\la ),
\end{eqnarray}
where
\begin{eqnarray}\label{multiener}
R(\la ) &=& {1\over 2\pi} \int e^{-i\omega \la}\big(-1 + \cr
&& \hskip .1in {1\over 2}
{\sinh (\pi\omega/\mu)\sinh (\omega ) \over 
\cosh (\omega )\sinh (\omega(\pi/\mu-2s))\sinh(2s\omega )}\big)\cr
e_0 &=& {\Lambda_c \over 2}\tan^{-1}\big[{\cosh({\pi\la\over 2})
\over \sinh ({\pi f\over 2\mu})}]\cr
p_0 &=& {\Lambda_c \over 2}\tan^{-1}\big[{\sinh({\pi\la\over 2})
\over \cosh ({\pi f\over 2\mu})}].
\end{eqnarray}
These equations are derived assuming $\pi/\mu > 2s$.
They, as they should, reduce to those of the U(1) Thirring model in
the AF region when s=1/2.  $\mu$ and $f$ are given in terms of
the bare parameters $J_\perp$ and $J_\parallel$ as
follows:
\begin{eqnarray}
J_\parallel &=& -2\mu\cr
J_\perp &=& 4\mu e^{-f}\big[ {\sin (\mu s)\over \mu s}\big]^2.
\end{eqnarray}
These relations
are valid in the limit of large $f$,
a limit we would take in any case
in enforcing the scaling limit.

\begin{figure}[tbh]
\centerline{\psfig{figure=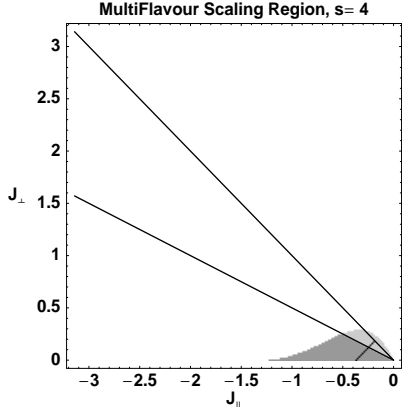,height=4.in,width=3in}}
\vskip -1in
\caption{The portion of the region of the multiflavour fermion model
that is well described by a field theory for s=4.   
The two diagonal lines delimit the region $J_\parallel > J_\perp/2$.
The more lightly shaded area
marks the intersection of the region satisfying
(\ref{multirel}) with the region given by 
$2J_\perp > J_\parallel > J_\perp/2$.}
\end{figure}

>From these relations, we again see that in order to maintain
a finite anisotropy ($\mu \neq 0$) while at the same time insisting
that the spectrum be purely relativistic, we must take $f \rightarrow \infty$.
Thus the scaling limit again requires $J_\perp \rightarrow 0$ and
thus an infinite bare anisotropy.  To keep 
a finite bare anisotropy, we
must take $J_\parallel \rightarrow 0$ simultaneously.  
Hence $\mu \rightarrow 0$
and again we obtain an isotropic theory.

As before, we now relax the strict scaling limit.
With $e_0$ and $p_0$ identical in form to the U(1) Thirring
case, we see again that we must satisfy
\begin{equation}\label{multirel}
{f\over \mu} > {1\over\pi}\log ( {4C^2A \over 3}),
\end{equation}
where again we have demanded that the spectrum appear relativistic
to one part in A up to scales of $C\times m$, where the fermion
mass scale is given by $m\sim \Lambda_c e^{-f/\mu}$.
The region of parameter space satisfying these constraints 
for $s=4$ and $A=C=50$
is plotted in Figure 9.  We 
see that the allowed region of bare parameters with a `physical'
anisotropy is small compared to the entire allowed parameter space.

As with the U(1) Thirring model, we ask how the magnetic susceptibility
depends upon the anisotropy.  Given the functional similarity of 
(\ref{multiener}) with (\ref{u1eq}), the susceptibility is given
(almost) identically by (\ref{afchi}):
\begin{eqnarray}\label{multichi}
\chi (H) &=& {1\over 2^{5/2} \pi} {1\over (1-2s\mu/\pi)^2}
\bigg\{ ({H-H_c \over H_c})^{-1/2}\cr 
&& - {2^{9/2} \over \pi}R(0) + {\cal O}({H-H_c \over H_c}) \bigg\},
\end{eqnarray}
where $R$ is now the kernel in (\ref{multiener}) and
\begin{equation}
H_c = 2m(1-2s\mu).
\end{equation}
In Figure 10 we plot $\chi (H)$ for values of the parameters 
lying along the small diagonal line shown in Figure 9.
We see, as with the U(1) Thirring AF region, the various plots
lie essentially on top of one another.  Again, the allowed
`physical' anisotropy does not lead to large variations in the
physics.

\begin{figure}[tbh]
\centerline{\hskip-.5in
\psfig{figure=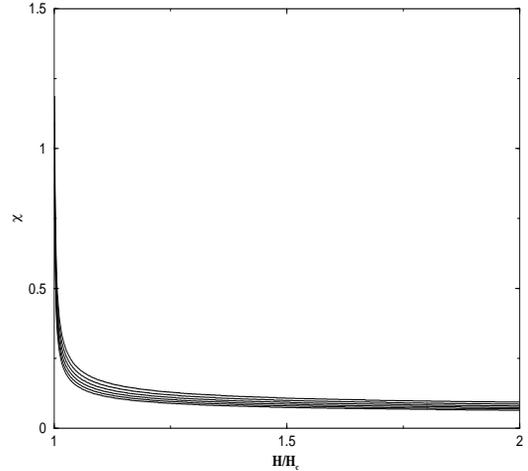,angle=-90,height=2.5in,width=2.5in}}
\caption{The dependence of the susceptibility upon
the anisotropy for $s=4$. From bottom to top, the plots correspond to
six equally spaced values of $(J_\parallel,J_\perp)$ along the
line $(-.2,2.)$ to $(-.4,0)$.}
\end{figure}

\subsection{Anisotropic principal chiral model $U(1)\times SU(2)$}

A model closely related to 
the previous one is the anisotropic SU(2) principal chiral model (APCM),
with 
the action
\begin{eqnarray}
S_{\rm APCM} &=& \int d^2x 
[{1\over J_\perp} ((\omega^x_\mu)^2 + (\omega^y_\mu)^2) +
{1\over J_\parallel} (\omega^z_\mu)^2]; \cr
\omega^a_\mu &=& {\rm Tr} (\sigma^a g^{-1} \del_\mu g),
\end{eqnarray}
where $g$ are matrices in the fundamental representation
of SU(2).

\begin{figure}[tbh]
{\centerline{\psfig{figure=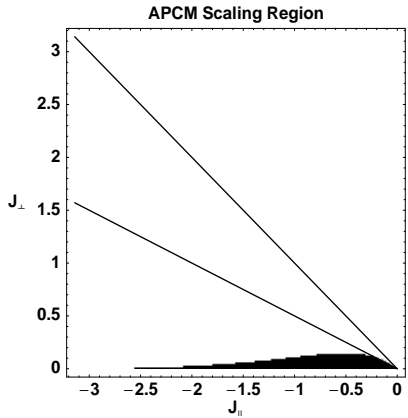,height=4.in,width=3in}}
\vskip -1in
\caption{The portion of the region of the anisotropic
principal chiral model 
that is well described by a field theory.
The two lines mark the region characterized by the condition
that $J_\parallel > J_\perp/2$.}}
\end{figure}

The APCM is intimately related
to models of N fermions as $N\rightarrow \infty$ \cite{Wie}.
In particular, the multifermion model of the previous section
was shown to be equivalent to the APCM when $s\rightarrow\infty$
provided $\mu \rightarrow 0$ (as the Bethe ansatz solution is
periodic in $\mu s$) while $\pi /\mu - 2s$ is held constant.
In this limit the bare parameters of the APCM model, $J_\perp$ and
$J_\parallel$, are given in terms of the parameters $\mu$ and $f$
by
\begin{eqnarray}
J_\parallel &=& -2\mu\cr
J_\perp &=& {16\mu \over \pi^2} e^{-f}.
\end{eqnarray}
Given the form of these equations, all the conclusions regarding
the scaling limit of the multiflavour fermion model hold.
In particular, we imagine holding s large and $\mu$ small but both
finite.  Then for the spectrum of excitations to be purely relativistic,
we must take the limit $f\rightarrow\infty$ 
thus implying $J_\perp \rightarrow 0$.  
To then avoid an infinite bare anisotropy, we have 
$J_\parallel\rightarrow 0$ and we end up with an isotropic theory.
If we relax the scaling limit to the same degree done for
the multiflavour model, we again find a finite region of
parameter space admitting a relativistic sector.  This
region, determined by (\ref{multirel}) with $A=C=50$, is 
plotted in Figure 11.  Unlike the multiflavour model
for finite s, this region does not see a ratio, $J_\perp/J_\parallel$,
of the bare parameters larger than $1/2$.

\subsection{The O(3) Sigma Model and the sausage model at $\theta=0$}

An amusing illustration of the 
interplay between anisotropy in the bare theory and in the 
scaling limit is
provided by 
the so called sausage model \cite{FZO},
an anisotropic deformation of the $O(3)$ sigma model. 
Setting $\phi_i\phi_i=1$, the 
action of the usual $O(3)$ sigma model reads
\begin{equation}
S={1 \over 2 g}\int \left(\partial_\mu\phi_i\right)^2,
\end{equation}
where $g$ is a running coupling constant given at 1-loop by
\begin{equation}
g = -{2\pi \over t},
\end{equation}
and $t$ is the `RG time'.  In terms of the cutoff, $t=\ln a$, and so
$t\to -\infty$ in the limit when
the cutoff is sent to zero. 
The S matrix 
describing this model has of course an $O(3)$ symmetry. 
There exists, as usual, an anisotropic  
deformation of this S matrix, which is characterized 
in \cite{FZO} by a parameter $\nu$ which in turn determines
a quantum
group parameter, $q\approx -e^{-i\nu\over 4}$ for small $\nu$. 
As argued in \cite{FZO}, the one loop action for this model then reads
\begin{equation}
S={1\over 2 g(t)}\int d^2x{\left(\partial_\mu\phi_i\right)^2\over 
1-{\nu^2\over 2g}\phi_3^2},
\end{equation}
where $g(t)={\nu\over 2}\coth{\nu (t_0-t)\over 4\pi}$. This action 
also corresponds to 
a sigma model, but in this case the  target space has the
shape of a ``sausage''. 
In order for this action to reproduce the physics contained in the
corresponding S-matrix, the authors of \cite{FZO} found it necessary
to take $t \to -\infty$ or equivalently, the cutoff, $a$, to 0.  
In this limit,  the sausage becomes a very long cylinder of
length $L={\sqrt{2\nu}\over 2\pi}(t_0-t)$, and 
circumference, $l=2\pi\sqrt{2\over\nu}$.  
For any $\nu\neq 0$, it thus follows that the asymptotic UV shape of 
the sausage as $a\to 0$ is 
infinitely elongated, i.e. $L/l\to\infty$.
Once again, a finite anisotropy, $\nu \neq 0$, in the scaling limit
requires an infinite bare anisotropy.

The sausage model
in fact bears a close resemblance to the anisotropic Kondo 
and $U(1)$ Thirring model. To see this,
one writes the one loop action using stereographic coordinates
\begin{equation}
S=\int d^2x{\left(\partial_\mu X\right)^2 
+ \left(\partial_\mu Y\right)^2 \over a(t)+b(t)\cosh 2Y}, 
\end{equation}
and shows that the standard RG equations for 
the metric \cite{friedan} give rise to 
\begin{eqnarray}
{da\over dt}={1\over 2\pi} b^2\nonumber\\
{db\over dt}={1\over 2\pi} ab,
\end{eqnarray}
with the invariant $\nu^2=a^2-b^2$. We have not worked out in detail the 
physical properties of this model, but the similarity of the equations
with those of the anisotropic Kondo and $U(1)$ Thirring model 
(with $\mu$ identified with $\nu$) suggest that the qualitative results 
in sections A and B above will hold here as well.

\section{Lessons from the Scaling Limit: Restoration of General Symmetries
Through the RG}

We now apply the lessons drawn in the previous section to 
a wider class of theories.  Specifically, we consider current-current
perturbations of Wess-Zumino-Witten (WZW) models together with a set
of Kondo problems where the underlying symmetry is 
$SU(2)_k$.  We show that the parameter
spaces of these models possess subspaces of higher symmetry to which
the 1-loop RG flows.  Given the scaling limit implies that one
can only realize a field theory if the bare couplings are weak 
(unless one is ready to accept unphysically large bare anisotropies), we
argue that these 1-loop flows are to be trusted, subject to certain
caveats.
To begin, we consider current-current perturbations of WZW models.

\subsection{Current-Current Perturbations of WZW models} 

Here we consider a set of theories represented by
current-current perturbations of 
WZW models based upon simple groups, $G$:
\begin{equation}\label{i}
S = S_{WZW}^G + \sum_{a=1}^n g_a \int d^2x J_a^L\cdot J_a^R ,
\end{equation}
where n is the dimension of the group.
The 1-loop beta function of these theories has the form:
\begin{equation}\label{iii}
{d g_a \over dl} = - \sum_{bc} (f_{abc})^2 g_b g_c ,
\end{equation}
where $f_{abc}$ are the structure constants for $G$.
\footnote{The structure constants are generated by 
some matrix representation, $t^a$ via $[t^a,t^b] = if^{abc}t^c$.
We suppose the $t^a$'s are such that $Tr(t^at^b) = 2\delta^{ab}$.}
Such $\beta$-functions are predicated upon the second term of the 
current-current operator product expansion (OPE):
\begin{equation}\label{iv}
J_a (x) J_b (y) = {k\over (x-y)^2}\delta_{ab}+{1\over (x-y)}i f_{abc} J_c (y),
\end{equation}
where $k$ is the level of the WZW model.  
The above $\beta$-functions are level independent, a consequence
of the second term not depending upon k.

It is straightforward to show that there are regions of parameter space
where (\ref{iii}) implies a restoration of symmetry, i.e. the RG flows
onto a ray described by $g_a = g$, $\forall a$. 
We clearly see that the ray itself
is an RG invariant.  Setting $g = g_a$ on the r.h.s. of (\ref{iii}),
we have
\begin{equation}\label{v}
{d g_a \over dl} = - g^2 \sum_{bc} (f_{abc})^2 = -c_v g^2 ,
\end{equation}
where $c_v$ is the quadratic casimir of the adjoint representation
for the group $G$.
The question then becomes whether the ray actually attracts the flow. 

To answer this we perform a linear stability analysis.
Writing
\begin{equation}\label{vi}
g_a = g + \delta g_a ,
\end{equation}
we obtain RG equations for $\delta g_a$:
\begin{eqnarray}\label{vii}
{d \delta g_a \over dl} 
&=& - g \sum_{bc} (f_{abc})^2 (\delta g_b + \delta g_c )\cr
&=& -2g \sum_{bc} (f_{abc})^2 \delta g_b .
\end{eqnarray}
Thus all we need to do is find the eigenvalues/eigenvectors of the above
equation.  We see that the eigenvector $\delta g_a = 1 $
(i.e. directed along the isotropic ray) has an eigenvalue, $\lambda = -2gc_v$.
If $g <0$, we claim that it is the largest positive 
eigenvalue.  To see this let
\begin{equation}\label{viii}
v = (\alpha_1, \ldots ,\alpha_n), 
\end{equation}
be an eigenvector of (\ref{vii}) with eigenvalue, $\lambda_v$,
where the couplings $\alpha_i$
are chosen without loss of generality
such that $1\geq |\alpha_1| \geq |\alpha_2| \geq \ldots \geq |\alpha_n|$.
Then the eigenvalue $\lambda_v$ is constrained by
\begin{eqnarray}\label{ix}
|\lambda_v \alpha_1| &=& |2 g \sum_{bc} (f_{1bc})^2 \alpha_b|\cr
&\leq & |2g|\sum_{bc} (f_{1bc})^2 |\alpha_b|\cr
&\leq & |2g|\sum_{bc} (f_{1bc})^2 |\alpha_1|\cr
&=& |2g||\alpha_1| c_v.
\end{eqnarray}
Thus $\lambda_v \leq 2|g|c_v$, and the isotropic ray is the most
relevant direction.

Although we have established that the most relevant RG direction
for $g < 0$ is the isotropic ray, we have not dealt with the
existence of other less (but still) relevant directions in the RG, i.e.
beyond $-2gc_v$ there are still a set of positive eigenvalues
to the linear analysis of (\ref{vii}).  Unlike the eigenvalue
corresponding to the isotropic ray, these eigenvalues are non-universal in
the sense that they depend upon the particular form of the structure
constants.  Nevertheless it is instructive to compute the value of the
next largest eigenvalue for the various groups.
\footnote{To compute the relevant structure constants for the various
groups, we employ the following conventions for the generators, $t^a$.
Define a set $N\times N$ matrices, $e^N_{pq}$, such that 
$(e^N_{pq})_{ij} = \delta_{ip}\delta_{jq}$.  For the SO(N) series, we
take the generators to be of the form $(ie^N_{pq}-ie^N_{qp})$, $p < q$,
for a total of $N(N-1)/2$ generators.  For the SU(N) series we 
choose the generators to be of the form, 
$b_{pq} = (e^N_{pq}+e^N_{qp})$, $p < q$,
$c_{pq} = (ie^N_{pq}-ie^N_{qp})$, $p < q$, and 
$h_{mm} = ((m+m^2)/2)^{-1/2}(\sum^m_{j=1}e^N_{jj} - me^N_{mm})$, 
$1\leq m < N-1$
for a total of $N^2 -1$ generators.
And finally for the SP(2n) series, the generators have the
form $\big( {X_1~~X_2 \atop X_2^\dagger~~-X_1^T}\big)$ with 
$X_i$ being $N\times N$ matrices, $X_1$ hermitian, and $X_2$ symmetric. 
As such $N^2$ SP(2n) generators can be obtained by choosing $X_2 = 0$ and
$X_1$ 
to be $b_{pq}/\sqrt{2}$, $c_{pq}/\sqrt{2}$, or $e^N_{pp}$, $1\leq p \leq N$.  
We obtain another
$N(N-1)$ generators by choosing $X_1 = 0$ and 
$X_2 = (e^{N}_{pq}+e^{N}_{qp})/\sqrt{2}$
or $X_2 = i(e^{N}_{pq}+e^{N}_{qp})/\sqrt{2}$, $p<q$.  
The final 2N SP(2N) generators
are arrived at via $X_1 = 0$ and $X_2 = e^N_{pp}$ or $ie^N_{pp}$.}
The results are in the table below:

\vskip 10pt
\centerline{\begin{tabular}{|c|c|}\hline
$G$ &  $\gamma = \lambda /|2gc_v|$\\
\hline\hline
$SO(N\geq 5)$ & $(N-4)/(2N-4)$\\
\hline
$SU(N \geq 3)$ &  $1/2$\\
\hline
$Sp(2N)$ & $(2+N)/(2+2N)$\\
\hline
\end{tabular}}
\vskip 10pt

\noindent Here the second 
column gives the ratio of the next largest eigenvalue
to that of the largest, $2|g|c_v$, for all the non-exceptional groups.
Although these ratios are not universal, it is worthwhile 
to note that none are close to one and so no other ray approaches
the isotropic direction in importance.

As discussed by \cite{Lin} in the context of a current-current perturbation
of an SO(8) level 1 WZW model, i.e. an anisotropic SO(8) Gross-Neveu
model, the presence of these less relevant directions
does not destroy the symmetry restoration.  Rather it introduces corrections
to the theory
that behave as the power $1-\gamma$ of the bare coupling.  
As $\gamma$ is not close to unity, small bare 
couplings induce small corrections.  We, following \cite{Lin}, can make 
this statement precise.

Let $g_i$ be the coupling along the isotropic ray and let
$g_b$ be the coupling along the next most relevant direction breaking
the symmetry.  We 
determine the $\beta$-functions of these couplings from the 
$\beta$-functions of the linear stability analysis:
\begin{eqnarray}\label{x}
{dg_i \over dl } &=& -g_i^2;\cr
{dg_b \over dl } &=& -\gamma g_i g_b; ~~~ \gamma < 1 ,
\end{eqnarray}
where $\gamma$ is given in the above table.  Integrating these
equations we find
\begin{eqnarray}\label{xi}
g_i (l) &=& {g_i (0) \over 1 + g_i (0) l}\cr
g_b (l) &=& {g_b (0) \over (1 + g_i (0) l)^\gamma},
\end{eqnarray}
where $g(0)$ marks out the bare value of the couplings.

In order to ascertain the effect of $g_b$ on the symmetry
restoration, we consider its effect on the gaps/masses of the
model.  To determine the gaps we use the relation relating the
bare/physical gaps to the renormalized gaps:
\begin{equation}\label{xii}
\Delta (g_i(0) , g_b (0)) = e^{-l} \Delta (g_i (l),g_b (l)).
\end{equation}
Typically the above equation is evaluated at $l = l_c$ where $l_c$, the cutoff
scale, is determined by $g_i (l_c) = 1$, the point where the 1-loop
RG breaks down.  From (\ref{xi}), $l_c$ is given by
\begin{equation}\label{xiii}
l_c = -{1\over g_i(0)} - 1.
\end{equation}
To determine the effect of $g_b$ on the gaps we consider the
ratio of two gaps:
\begin{equation}\label{xiv}
{\Delta_1 (g_i(0),g_b(0)) \over \Delta_2 (g_i(0),g_b(0)) }
= {\Delta_1 (1,g_b(l_c)) \over \Delta_2 (1,g_b(l_c)) }.
\end{equation}
As $\Delta_1 (1,0) = \Delta_2 (1,0)$ by the underlying symmetry, this
ratio reduces to 
\begin{equation}\label{xv}
{\Delta_1 (g_i(0),g_b(0)) \over \Delta_2 (g_i(0),g_b(0)) }
= 1 + c g_b(l_c) ,
\end{equation}
where c is some number.  Using (\ref{xi}) and (\ref{xiii}) we then
have
\begin{equation}\label{xvi}
{\Delta_1 (g_i(0),g_b(0)) \over \Delta_2 (g_i(0),g_b(0)) }
= 1 + c g_b(0)g_i(0)^{-\gamma} .
\end{equation}
Then as $g_i,g_b \rightarrow 0$, the symmetry restoration becomes
exact.  With small but finite couplings the ratio of two gaps
is $1 + (U/t)^{1-\gamma}$, where $U$ is the typical bare coupling strength
and $t$ is the bandwidth.

Thus we have established that the $\beta$-functions of (\ref{i}) indicate
a symmetry restoration in a portion of weak coupling parameter space.  However
this symmetry restoration might in general be much wider.
In the work of \cite{fisher}, where the previously mentioned
example of a restoration
to an $SO(8)$ symmetry in the context of two-leg Hubbard
ladders was studied in great detail, it was found that the entire
parameter space at weak coupling saw a restoration of symmetry.
They found additional rays characterized by $|g_a| = g$ but with
sign variations among the $g_a$.  These additional rays marked
out different phases of the Hubbard ladders.  Included among
the phases were a D-Mott phase, a phase characterized by 
an interaction induced charge gap with D-wave symmetry, an S-Mott
phase, a CDW phase, and a spin-Peierls phase.
Presumably such additional rays
are present generically, but lacking physical
motivation, we do not search for them here.

Having established the $\beta$-functions in (\ref{iii}) imply a restoration
of symmetry, we ask if this RG-induced restoration is actually realized.  
Although it is impossible to answer this definitively, we can examine
this question in the light of the caveats concerning symmetry restoration
through an RG.
The first of these concerned the presence of an RG invariant that controls
the physics.
As the RG flow leaves the invariant unchanged,
we cannot expect the physics along the flow to become
more isotropic.  However we conjecture that the presence of 
this physics-controlling RG invariant is related to the presence of 
an additional symmetry in the theory.
In the case of the U(1) Thirring model, we found that the
RG invariant, $\mu$, which governed the physics, parameterized
the deformed quantum group
symmetry, $\widehat{sl(2)}_q$.
In the U(1)
Thirring case, this quantum group symmetry arose as a deformation of an
sl(2) Yangian symmetry present in the isotropic theory.  However the
U(1) Thirring model is somewhat unique in this regard.  
The isotropic current-current perturbations of level 1 WZW models all
possess similar Yangians.  But there is no sensible way in which
they can be deformed.
There do
exist generalizations of the U(1) Thirring model which are characterized
by deformed quantum group symmetries\cite{bazhanov}.  However these models
correspond to imaginary coupling Toda theories and on the face of it
are not even hermitian.  Thus for symmetries larger than SU(2),
we do not expect to be faced with this particular problem.  Of
course, it is conceivable that other hidden symmetries lurk in
these models.

We must also be aware of the possibility that the IR fixed point to which
the RG flows is not so much a point but a ray, that is, there exists a
marginal operator at the fixed point such as in the deformed O(3) sigma 
model with topological angle $\theta = \pi$.  This is obviously not a
concern when the fixed point is massive and so is unlikely to be
relevant in the cases of bulk current-current perturbations of WZW models.

\subsection{$SU(2)_k$ Overscreened  Kondo Problems}

In this subsection we consider a set of
overscreened
$SU(2)$, k-channel Kondo models.
Such models arise
when 
k channels of spin-1/2 fermions are coupled to an
impurity spin, $s^a$, of magnitude, $s<k/2$.
These theories can be represented\cite{affleck}
as boundary perturbations of
a chiral $SU(2)_k$ WZW model  (with Hamiltonian
$H^{L}_{WZW}$):
\begin{equation}
\label{xvii}
H = H^{L}_{WZW}  + \sum_{a=x,y,z} \  g_a \   J_L^a(0)\cdot s^a.
\end{equation}
As usual
we  have represented these Kondo problems in their 
unfolded realization: only one chiral ($L$-)
component of the local spin density
 of the bulk theory, $J^L_a(0)$, 
couples to the impurity spin, $s^a$.  

In the following we restrict our discussion of these anisotropic models to
an  $s=1/2$ impurity spin. Indeed, in this case these models
share the same 1-loop beta functions as the current-current
perturbations of bulk WZW models, considered in the previous subsection.
Therefore, the overscreened Kondo models also share the
same putative RG-induced symmetry restoration with 
the current-current perturbations. The question then
becomes whether this symmetry restoration suggested
by the 1-loop RG equations is genuine.

The answer to this question is affirmative:
this follows from the analysis of the isotropic
overscreened strong coupling fixed point, done in 
\cite{cox}, where the latter was found to be stable
with respect to small spin-exchange anisotropic 
perturbations\footnote{Recall that we are discussing the
case of an $s=1/2$ impurity spin.}.  In other words,
isotropy in the spin-exchange couplings is 
restored, and the conclusions obtained from
the 1-loop RG analysis can indeed be trusted.

When the number of channels is large,  $k >> s=1/2$,  this can be
easily understood on the basis of the 2-loop RG equation
for the isotropic model (i.e. $g = g_a$), which has
the form \cite{noz&bla,affleck} 
\begin{equation}\label{xviii}
{d g \over dl} = - {c_v \over 2}g^2 - {k c_v \over 4} g^3 + ...,
\end{equation}
where here $c_v$, the casimir of the adjoint representation, is $2$.
We see that the number, $k$, of channels
 appears in the two-loop term.  This two
loop $\beta$-function thus implies a fixed point at 
$g_* \sim {1 \over k}$.
For large $k$ this fixed point appears at small coupling, in the range
of validity of the perturbative RG.  Moreover it is known 
\cite{affleck,noz&bla}
that higher loop terms do not destroy this fixed point provided k is
taken to be large compared to $3/4$, 
the value of the quadratic casimir of the 
representation of spin-1/2.

To exploit this behaviour in the isotropic $\beta$-function, we suppose
that the initial bare anisotropic couplings are much
smaller than the isotropic fixed point at ${\cal O} ({1\over k})$.
In this regime the 1-loop term in the $\beta$-function controls
the flow and symmetry is thus restored.  The theory will then proceed
to the isotropic fixed point, undisturbed by higher loop terms, 
as in the purely isotropic case.

We would like to stress that these issues of symmetry restoration
become more
involved for the RG flows in the  overscreened multi-channel Kondo 
models with impurity spin $s>1/2$.
In all those cases (except when $k/2 = (s-1/2)$ or $k\leq 4$), the
isotropic overscreened strong coupling fixed was found unstable
to small anisotropies in the spin-exchange couplings\cite{cox}.
At first sight the anisotropic 1-loop RG equations appear to be 
the same as in
the case of impurity, spin $s=1/2$.  Thus as these predict
symmetry restoration, 
one may then be tempted to conclude that the 1-loop RG in this case
is unreliable.

However, a careful look at the 1-loop RG
with exchange anisotropic couplings reveals that
new terms of zero
dimension in the Hamiltonian of
the form,
\begin{equation}
\label{SaSa}
h_a  S^a S^a ,
\end{equation}
will be generated.
(Note that for spin, $s=1/2$, such terms are proportional to the
identity and so serve to only renormalize the impurity free energy.)
Since these terms are highly relevant, they
have a profound effect on the
RG flows, which are therefore very different from
the case of impurity spin, $s=1/2$. 
Indeed, in the large-$k$ limit it is those
dimension zero
(bare) operators which renormalize into the
relevant symmetry breaking perturbation
of scaling dimension, $\Delta=6/(2+k) \to 0$, found
in \cite{cox}.

\section{Symmetry Restoration in a broken U(1) Thirring Model}

In the previous sections, we have argued that while the U(1) Thirring
model sees
symmetry restoration as
more broadly understood in
the course of taking the scaling limit, it does not experience
an RG-induced
restoration of symmetry.
This occurs in the $U(1)$ Thirring model because physical quantities
depend upon the RG invariant, $\mu$.
The quantity $\mu$ is a reflection of the
quantum group symmetry $\widehat{sl(2)}_q$ present in U(1) Thirring, and 
it is  natural to determine the consequences of breaking
this symmetry.  This does not necessarily mean that in breaking
the symmetry, we exclude the possibility of a $\mu$-like parameter.
$\mu$ is an RG invariant.  In breaking the symmetry, one does not
eliminate all RG invariants.  One merely alters their form.
Indeed as RG trajectories are lines, they can generally be
parametrized by  quantities 
that will be constant along the RG flow, say the intercept with one of the 
hyper-planes in the RG space. 
What is not clear
is to what extent the physical properties are going to depend on 
that parameter. In breaking the symmetry, we will in general
lose any known means to connect
the unchanging physical properties with the RG invariant.
It is akin
to finding accidental degeneracies in QM.  Such accidental degeneracies
are almost always related to hidden symmetries.  With no symmetries,
there are no such degeneracies.  It is thus tempting to propose that with no
symmetries around to be associated with the RG invariants, these invariants
cannot, in general,  govern the physics, and therefore that an RG-induced
symmetry restoration should generally occur.

A simple example where this idea can be investigated is a broken
U(1) Thirring model.  We thus consider what we call the $g_{xy}$-Thirring
model:
\begin{eqnarray}\label{gxymodel}
{\cal L} = i\bar\psi_\alpha\gamma_u\del^\mu\psi_\alpha + 
{1\over 4}g_\parallel (j_z)^2 &+& {1\over 4}g_\perp((j_x)^2 + (j_y)^2)\cr
&+& {1\over 4}g_{xy} j_xj_y .
\end{eqnarray}
The one loop RG equations for this model are
\begin{eqnarray}
{dg_\parallel \over dl} &=& - g^2_\perp + {g_{xy}^2\over 2};\cr
{dg_\perp \over dl} &=& - g_\perp g_\parallel ;\cr
{dg_{xy} \over dl} &=& {g_{xy} g_\parallel\over 2} .
\end{eqnarray}
Recalling that $\mu^2 = g_\parallel^2 - g_\perp^2$, we see that
the $g_{xy}$-perturbation causes $\mu^2$ to flow:
\begin{equation}
{d\mu^2 \over dl} = g_\parallel g_{xy}^2.
\end{equation}
In the AF region, $g_\parallel$ is negative and so $\mu^2$ decreases
under the RG.  However, as $g_{xy}$ is (marginally) irrelevant in the AF
region, we do not necessarily expect a full restoration of the SU(2)
symmetry: as $g_{xy}$ goes to zero, $\mu^2$ will stop flowing.

In region C, $\mu^2$ is negative while $g_\parallel$ is either
positive or negative.  Thus depending on the bare value of $g_\parallel$,
$\mu^2$ either increases or decreases in magnitude.
However again 
$\mu^2$  will not change without bound as
$g_\parallel$ eventually flows to negative values.  Moreover a change 
of $\mu^2$ in region C corrects the physics only minimally
due to the exponentially small dependence of physical quantities
upon $\mu$ in this region.

Thus it would seem that in breaking $\widehat{sl(2)}_q$, there
is an RG-induced symmetry restoration in the model where there
was none before.
However we offer a cautionary tale on the interpretation of
the above analysis.
We now choose to modify the U(1) Thirring model by
instead introducing separate couplings for $j_x^2$ and $j_y^2$:
\begin{eqnarray}
\cal{L} &= i\bar\psi_\alpha\gamma_u\del^\mu\psi_\alpha + 
{1\over 4}g_\parallel (j_z)^2 + {1\over 4}g_x (j_x)^2 + {1\over 4}g_y(j_y)^2.
\end{eqnarray}
In this XYZ-Thirring model, 
the one loop RG equations become
\begin{eqnarray}
{dg_\parallel \over dl} &=& - g_x g_y ;\cr
{dg_x \over dl} &=& - g_y g_\parallel ;\cr
{dg_y \over dl} &=& - g_x g_\parallel .
\end{eqnarray}
Unlike (\ref{gxymodel}), this model is known to be
solvable by Bethe 
ansatz.  The results are similar to those 
of the $U(1)$ Thirring model, trigonometric functions being replaced
by elliptic functions.  Following \cite{Dutyshev} and \cite{DestriDeVega},
one finds for instance the dispersion relations
\begin{eqnarray}
e_0/p_0(\lambda )={\Lambda_c\over 4}\left\{
 \tan^{-1}\left[{\hbox{cn}(\tilde{f}-{\pi\la\over 2},\tilde{k})\over 
\hbox{sn}(\tilde{f}-{\pi\la\over 2},\tilde{k})}\right]\pm\right. \nonumber\\
 \left.\tan^{-1}\left[{\hbox{cn}(\tilde{f}+{\pi\la\over 2},\tilde{k})\over 
\hbox{sn}(\tilde{f}+{\pi\la\over 2},\tilde{k})}\right]\right\}.
\end{eqnarray}
Here $cn$, $sn$ and $dn$ are the usual elliptic functions.
The parameter $\tilde{f}$ is defined by $\tilde{f}={f\over\mu}K'(\tilde{k})$, 
where in turn the dual modulus $\tilde{k}$ is defined by 
$K'(\tilde{k})K'(k)=\mu K(\tilde{k})$ with $k$ being the original modulus.
The key parameters $\mu,k,$ and $f$ are 
obtained in terms of the bare coupling constants as
\begin{eqnarray}\label{xyzrel}
k \hbox{sn}^2 (\mu,k) &=& \tan\left({g_x-g_y\over 2}\right) 
\tan\left({g_x+g_y\over 2}\right)\nonumber ;\\
\hbox{cn}(\mu,k)\hbox{dn}(\mu,k) &=& {\cos g_\parallel\over 
\cos\left({g_x-g_y\over 2}\right)
\cos\left({g_x+g_y\over 2}\right)}; \nonumber\\
{\hbox{sn}(if,k)\over \hbox{sn}(\mu-if,k)} &=&
-{\cos\left({g_x+g_y\over 2}\right)
\over \cos\left({g_x-g_y\over 2}\right)} e^{-ig_\parallel}.
\end{eqnarray}
Both $\mu$ and $k$ are RG invariants, as readily can be checked.
All physical quantities are determined in terms of these parameters
and so in the XYZ-Thirring model 
there is no RG induced symmetry restoration.

The absence of symmetry restoration
does not {\it necessarily} contradict our notions on accidental degeneracies.  
Here we
have not actually broken the $\widehat{sl(2)}_q$ symmetry.  Rather
we have only deformed it into its elliptical cousin \cite{Sklyanin}.
Unfortunately, it is easily
possible to convince oneself otherwise.
Defining $\delta g_\perp = g_x - g_y \ll g_x,g_y$ and 
$g^{av}_\perp = {1\over 2}(g_x + g_y)$, the former as a measure 
of how broken the U(1) symmetry is,
we find the following RG equations:
\begin{eqnarray}\label{xyzrg}
{d\delta g_\perp \over dl} &=&  g_\parallel \delta g_\perp ;\cr
{dg_\parallel \over dl} &=& {1\over 4}(\delta g_\perp)^2 - (g^{av}_\perp)^2 
;\cr
{dg^{av}_\perp \over dl} &=& - g_\parallel g^{av}_\perp .
\end{eqnarray}
In region AF we have $g_\parallel < 0$.  If we did not know the
model was integrable, it would be natural to define the 
analog to $\mu$ in the broken $U(1)$ case to be
\begin{equation}\label{def}
\mu^2 = {g_\parallel}^2 - (g^{av}_\perp)^2 .
\end{equation}
We then see
\begin{equation}\label{murg}
{d\mu^2 \over dl} = {g_\parallel \over 2}(\delta g_\perp)^2 .
\end{equation}
Hence under the RG, $\mu^2$ decreases.  Thus it would seem that in breaking
the $U(1)$, 
an additional increase in symmetry is achieved.  However this does
not mean the full SU(2) symmetry is restored, as with $g_\parallel < 0$,
we see the quantity driving the flow of $\mu$, $\delta g_\perp$,
is itself in fact irrelevant.  Thus we have a conclusion similar
in spirit to the $g_{xy}$-Thirring model.
In region C, we again see similarities between the XYZ and 
$g_{xy}$-Thirring models.  In this region, $\mu^2$ 
is negative while $g_\parallel$ is either
positive or negative.  So again $\mu^2 $ will either increase or
decrease, but not without bound.  And the apparent physical
consequence of this is minimal given the exponentially small dependence
of physical quantities upon $\mu$.

Thus given our definition of $\mu$ (\ref{def}), we see a 
putative partial restoration
of symmetry under the RG.  But as we know from the exact solution
this does not actually occur.  It is merely an artifact of our definition
of $\mu$.  Thus in saying
a (partial) RG-induced symmetry restoration occurs in the $g_{xy}$-Thirring
model, we are implicitly supposing that this XYZ-Thirring scenario
is not applicable.  In particular, we are supposing 
that the $g_{xy}$-perturbation has 
genuinely broken the $\widehat{sl(2)}_q$
symmetry and not merely deformed it, thus forbidding an exact solution.

Although there is no RG-induced symmetry restoration 
of the XYZ-Thirring model, the above 1-loop RG analysis 
(\ref{xyzrg}-\ref{murg})
is reflected in the exact solution.  At weak coupling appropriate
to the 1-loop analysis, one must take the modulus, $k$,
of the elliptic functions in the relations (\ref{xyzrel}) to zero,
leaving 
\begin{equation}
\cos\mu\approx {\cos 2g_\parallel\over \cos (g_x-g_y)\cos(g_x+g_y)}.
\end{equation}
For a fixed $g_\perp^{av}$, increasing the measure of the $U(1)$ breaking,
$\delta g_\perp$ leads to smaller values of $\mu$, 
confirming the one loop RG results. 

The 1-loop RG is also reflected in a more elemental analysis of the XYZ
problem.
For simplicity consider an XYZ Kondo model instead, with couplings 
$g_x , g_y$, and $g_\parallel$.  
After bosonization, and forgetting inessential constant 
factors, the Hamiltonian reads
\begin{eqnarray}
H &=& H_0 + (g_x+g_y)\left[s^+ e^{i\sqrt{8\pi}\phi(0)}+s^- 
e^{-i\sqrt{8\pi}\phi(0)}\right] \cr
&&\hskip -.25in \!+(g_x-g_y)\left[s^+e^{-i\sqrt{8\pi}\phi(0)}
+s^-e^{i\sqrt{8\pi}\phi(0)}\right]\!+\!g_\parallel \partial_x\phi .
\end{eqnarray}
Suppose we now perform the canonical transformation, $U$, again
(see text before (7)). 
This time, the vertex operators involved in 
the second part of the Hamiltonian (that 
break the $U(1)$ symmetry) see their dimension {\sl increased}: 
they thus become irrelevant, and disappear in the scaling limit. 
Thus the 1-loop RG seems to correctly predict that the symmetry
breaking term, $g_x\neq g_y$, vanishes in the low energy limit.
Of course, bosonization
combined with a  canonical transformation
gives similar results for the XYZ Thirring model.

\section{Conclusion} 

This article has two overarching themes, both interconnected
through the scaling limit.  Firstly, we expanded
the notion of symmetry restoration to include all situations
in which the physical properties of a model have a weak dependence
upon an anisotropy.  When this definition was combined with constraints
coming from taking a scaling limit, we found (as discussed in
Section II)
that a wide variety of
models in fact see symmetry restoration.  Secondly, we observed 
that the scaling limit in general restricts the range of bare parameters
to be small.  We then exploited this fact in 
Section III
to argue that a 1-loop RG should accurately describe the physics.

Perhaps an underemphasized result of this work is that the
trustworthiness of the 1-loop RG has been previously underestimated.  
Even in cases where it seemed
to promise symmetry restoration which in fact did not occur (the
U(1) Thirring model), the fault lay not in the 1-loop RG but in its
interpretation.  Indeed the 1-loop RG of the U(1) Thirring model correctly
predicts the parameter $\mu$ to be an RG-invariant.  The problem, in contrast,
was in assuming that the physics will be governed by a ratio of
couplings, $g_\parallel/g_\perp$, and not $\mu$.  When the 1-loop RG
does indeed fail, for example in the $O(3)$ sigma model with $\theta = \pi$,
the reason is readily apparent: the topological term is manifestly
non-perturbative.

On a finishing note, the reader should certainly
not be left with the impression that symmetry restoration
always occurs in all possible models. 
There are many situations where it does not.
In many cases
this is again faithfully represented by the one loop RG. 

Indeed, in all the examples discussed in the bulk of the paper, the 
IR strong coupling fixed point was ``truly isotropic'', 
in the sense that it  
was $\mu$-independent. This is trivial for the
massive cases where there are no
leftover {\it massless}
 degrees of freedom in the IR; this is also true for the 
spin $1/2$, 1-channel  Kondo problem
where the impurity
spin is entirely screened by the conduction 
electrons. In the latter case
the isotropy of the fixed point manifests itself 
technically in the independence of the 
impurity scattering (reflection) matrix upon
$\mu$.   The operators determining the approach to the fixed point
have  $\mu$-independent dimensions, and it is only their respective
amplitudes that depend upon the anisotropy.  

As an obvious candidate lacking symmetry restoration
according to either definition, consider
the channel-anisotropic multichannel Kondo problem with 
impurity spin, $s=1/2$.
This model possesses an anisotropic (massless) IR fixed point.
Here we find the exact opposite to 
what we have seen so far;
that is, instead of a symmetry restoration,
there is an enhancement in asymmetry. 
The corresponding Hamiltonian is given by:
\begin{equation}
H=H_0+2\sum_{a,b=1,2}\sum_{m=1}^f 
 \ J_m \  ( \Psi_{a,m}^\dagger 
{\vec \sigma}_{ab}\Psi_{b,m})(0)  \cdot  {\vec s}.
\end{equation}
Here $a,b$ is the  spin index, and $m$ the  channel index.
$H_0=-i\sum_{a,m}\int dx \Psi^\dagger_{a,m}(x)\partial_x\Psi_{a,m}(x)$
describes non-interacting partial-wave electrons. 

Consider the simplest case with
$f=2$ channels (generalizations to more channels, $f>2$, are
straightforward with the same conclusions).
The physics of the channel-anisotropic case is simple\cite{noz&bla}
(see also \cite{cox}): the more strongly coupled channel
undergoes an ordinary (1-channel) Kondo effect, screening
the spin-1/2 impurity, whereas the more weakly coupled channel
decouples from the impurity.
Thus, the IR behavior of the channel-anisotropic
case is completely different from the (overscreened) channel-isotropic
situation. Clearly, channel-isotropy is not restored,
according to either definition (1 or 2 of the introduction).

This feature is easily seen in the one loop RG equations, 
which read in this case,
\begin{eqnarray}
\label{channelanisoRG}
{d J_1\over dl}=-C J_1^2\nonumber;\cr
{d J_2\over dl}=-C J_2^2 .
\end{eqnarray}
Thus the ratio ${\cal A}\equiv {J_1\over J_2}$ obeys
\begin{equation}
{d{\cal A}\over dl}=C J_1 {\cal A}(1-{\cal A}),
\end{equation}
and so grows under the RG.  Observe also that at the channel-isotropic
low energy fixed point (the non Fermi liquid Kondo
fixed point)   channel anisotropy is strongly relevant\cite{cox}.
There are also known cases where the anisotropy is exactly marginal,
i.e. a finite bare anisotropy leads to a finite anisotropy in the scaling
limit \cite{SSi}.  

The channel anisotropic model is also integrable\cite{AndreiJerez}, 
allowing us to relate to our discussion of the scaling limit.
In the simplest case
of two channels, $f=2$, the Bethe ansatz solution 
leads to expressions 
for the two characteristic energy scales in the problem: 
$T_i\equiv D e^{-\pi\over J_1}$, $T_a\equiv D 
\cos\left({J_1\over J_2}{\pi\over 2}\right)
e^{-\pi\over J_2}$ ($J_1\leq J_2$), where $D$ is the bandwidth 
that has to be taken to infinity in the scaling limit.  The ratio
$\Delta\equiv {T_a\over T_i}$ is the physical measure of anisotropy, the 
channel isotropic case corresponding to $\Delta=0$. 

The scaling limit is obtained by letting $D\to\infty$, $J_i\to 0$.  One 
checks then that keeping a ratio ${J_1\over J_2}$ finite,
i.e. a finite bare channel anisotropy, leads to $\Delta=\infty$, i.e. 
an infinite channel anisotropy in the scaling limit. 
The only way to have a finite anisotropy in the scaling limit is to start 
with an infinitesimally small bare anisotropy,
explicitly setting 
$J_1\approx {\pi\epsilon\over \ln (2\Delta/\pi\epsilon)}$, 
$J_2=J_1/(1-\epsilon)$, as 
$\epsilon\to 0$. The situation is thus the exact opposite of what 
we observed for the spin anisotropy: a finite channel anisotropy in the 
continuum limit 
requires an infinitesimally small bare channel anisotropy!

Another example of an anisotropic  (massless) IR fixed point  is
the spin $s=1$, $1$-channel Kondo problem.
The general
$s>1/2$, $k=1$ Kondo problem is
described by the 
Hamiltonian (\ref{firsteq}). 
Consider first the regime where
$g_\parallel > g_\perp  >0$.
Performing the same canonical transformation as in Section II.A leads
to a Hamiltonian as  in  (\ref{secondeq}).
This model is integrable if the 
impurity spin transforms in a spin $s$ representation of $sl(2)_q$, 
$q\approx -e^{-i\pi g_\parallel}$
(a quantum deformed version of $SU(2)$) \cite{fph}.
For $s=1/2$ or $s=1$ an ordinary spin can be used,
and no quantum deformation is necessary.
The resulting IR fixed point can be studied exactly.
For $s=1$ it is found to consist of a 
left-over conserved  spin $s'=1/2$.  The spin's
$z-$component induces on the electron degrees of freedom a phase
shift depending weakly upon the original anisotropy.
This is the usual anisotropic ferromagnetic 
Kondo effect, representing
a `line of anisotropic fixed points' continuously
deformable into the isotropic one.
Physical quantities at the fixed point,
such as the cross section for electron scattering of spin up or down, 
do depend upon the anisotropy, and so do the  operators
governing the approach to the fixed point. However
all these dependencies upon the anisotropy are weak (continuous),
and therefore we find symmetry restoration in the
expanded sense of definition 2.

When the size of the $sl(2)_{q}$ 
impurity is $s>1$, the left over impurity
spin obeys $sl(2)_{q'}$ commutation relations, with
$q' \approx q$ for small anisotropies, and the same conclusions
hold as for the $s=1$ result.

The above nature of the flow can also be understood
on the basis of the weak coupling RG.
Consider spin $s=1$.
As discussed earlier, 
a highly relevant zero dimensional operator,
$h s^z s^z$, as 
in \ref{SaSa} will be generated.  
When the renormalized  coupling  $h$ reaches 
a magnitude of order unity,
spin flip processes stop, and so does the RG flow.
For an anisotropy,
$\mu^2 =g^2_\parallel - g^2_\perp >0$, as above,
the induced coupling $h$ is negative,
and hence the impurity spin can be only
in two states, $s^z=\pm 1$,
at those scales.  At the same time,
the residual renormalized
coupling, $g_\parallel$,
amounts to a phase shift on the electrons, as in
the integrable formalism.
As $\mu^2 \to 0$, the phase shift becomes $\pi/2$, appropriate
for the underscreened isotropic IR fixed point.

In summary, the region,
$\mu^2 =g^2_\parallel - g^2_\perp \geq 0$, of the $s=1$
Kondo model flows 
into the ferromagnetic region
of the $s=1/2$ Kondo problem.
This flow can be represented through Figure 4.
Although Figure 4 is sketched for the conventions of the U(1)
Thirring model, by taking $g_\parallel \rightarrow -g_\parallel$ it can
be understood as that of the $s=1/2$ Kondo model.  Then
the $s=1$, $\mu^2 \geq 0$, Kondo model flows into region W of Figure 4.
Region AF is the usual anti-ferromagnetic Kondo regime.
Moreover, the boundary between regions C and W
describes the isotropic underscreened flow of the $s=1$ Kondo model.

There are however potential subtleties when $\mu^2 <0$.
A weak coupling analysis
including the zero dimensional operators as above indicates
that the region,
$\mu^2 =g^2_\parallel - g^2_\perp <0$, of the $s=1$ Kondo model
flows into region
C of Figure 4.  This is so since the induced coupling
constant $h$ is positive, freezing the impurity spin into the single
state, $s^z=0$.  Hence, the ultimate IR fixed point 
in this region would
consist of a  completely
eliminated impurity spin and no phase shift
for any non-vanishing amount of anisotropy,
in contrast to the underscreened isotropic
fixed point with phase shift, $\pi/2$.
Note that this lack of symmetry restoration
is already visible in the weak coupling RG.
This situation requires further study.

Another example that would be interesting to study in more detail
is the sausage model at $\theta=\pi$. 
In that case,
it is expected by analogy with the usual $O(3)$ model that the theory 
flows to a non trivial
fixed point at finite distance that is not accessible perturbatively
and is described by a 
compactified free boson with radius $R\approx 1+{\nu\over 8\pi}$ \cite{FZO}. 
No simple bare equivalent of this model
is known, and it is not clear that symmetry restoration in its expanded 
sense does occur: however, 
formal regularizations based on the Bethe ansatz equations certainly 
show a behaviour in all points identical with the previous cases.

\section{Acknowledgements}

The authors would like to acknowledge discussions with H.H. Lin.
R.K. has been supported by
NSERC, the University of Virginia, and the NSF both through grant
number DMR-9802813 and through the Waterman Award under
grant number DMR-9528578.  H.S. has been supported by
the Packard Foundation, the NYI
Program, and the DOE. 
H.S. also acknowledges hospitality and support from 
the LPTHE (Jussieu) and LPTMS (Orsay).

\end{document}